\documentclass{aa}

\usepackage{graphicx}
\usepackage{txfonts}
\usepackage{amsmath,amssymb}
\usepackage{soul}
\usepackage[dvipsnames]{xcolor}
\usepackage{soul}
\usepackage{caption} 
\usepackage{multirow}
\usepackage[colorlinks=true,citecolor=blue]{hyperref}

\newcommand{\itt}[1]{\textcolor{orange}{#1}}

\newcommand{\Euclid}{\textit{Euclid}}

\usepackage{etoolbox}
\makeatletter
\patchcmd\@combinedblfloats{\box\@outputbox}{\unvbox\@outputbox}{}{\errmessage{\noexpand patch failed}}
\makeatother
\begin{document}

\title{Validating the Fisher approach for stage IV spectroscopic surveys}
\titlerunning{Validating Fisher approach}
\author{
   S.~Yahia-Cherif\inst{1} \and 
   A.~Blanchard\inst{1} \and
   S.~Camera\inst{2,3,4} \and
   S.~Ili\'c\inst{5, 6, 1} \and
   K.~Markovi\v{c}\inst{7} \and
   A.~Pourtsidou\inst{8} \and 
   Z.~Sakr\inst{1,9} \and
   D.~Sapone\inst{10} \and 
   I.~Tutusaus\inst{11,12,1}
}

\institute{$^{1}$ Universit\'e de Toulouse, UPS-OMP, IRAP and CNRS,
IRAP, 14, avenue Edouard Belin, F-31400 Toulouse, France \\ \email{syahiacherif@irap.omp.eu}, \email{alain.blanchard@irap.omp.eu} \\
$^{2}$ Dipartimento di Fisica, Universit\`a degli Studi di Torino, Via P. Giuria 1, 10125 Torino, Italy\\
$^{3}$ INFN -- Istututo Nazionale di Fisica Nucleare, Sezione di Torino, Via P. Giuria 1, I-10125 Torino, Italy\\
$^{4}$ INAF -- Istituto Nazionale di Astrofisica, Osservatorio Astrofisico di Torino, Strada Osservatorio 20, 10025 Pino Torinese, Italy\\
$^{5}$ Universit\'e PSL, Observatoire de Paris, Sorbonne Universit\'e, CNRS, LERMA, F-75014, Paris, France \\
$^{6}$ CEICO, Institute of Physics of the Czech Academy of Sciences, Na Slovance 2, Praha 8 Czech Republic \\
$^{7}$  Jet Propulsion Laboratory, California Institute of Technology, 4800 Oak Grove Dr, Pasadena, CA 91109, USA\\
$^{8}$ School of Physics and Astronomy, Queen Mary University of London, Mile End Road, London E1 4NS, UK\\
$^{9}$ Universit\'e St Joseph; UR EGFEM, Faculty of Sciences, Beirut, Lebanon\\
$^{10}$ Departamento de F\'isica, FCFM, Universidad de Chile, Blanco Encalada 2008, Santiago, Chile\\
$^{11}$ Institute of Space Sciences (ICE, CSIC), Campus UAB, Carrer de Can Magrans, s/n, 08193 Barcelona, Spain\\
$^{12}$ Institut d’Estudis Espacials de Catalunya (IEEC), 08034 Barcelona, Spain
}
\authorrunning{S. Yahia-Cherif, A.Blanchard et al.}

\abstract{In recent years forecasting activities have become a very important tool for designing and optimising large scale structure surveys. To predict the performance of such surveys, the Fisher matrix formalism is frequently used as a fast and easy way to compute constraints on cosmological parameters. Among them lies the study of the properties of dark energy which is one of the main goals in modern cosmology. As so, a metric for the power of a survey to constrain dark energy is provided by the Figure of merit (FoM). This is defined as the inverse of the surface contour given by the joint variance of the dark energy equation of state parameters $\{w_0,w_a\}$ in the Chevallier-Polarski-Linder parameterisation, which can be evaluated from the covariance matrix of the parameters. 
This covariance matrix is obtained as the inverse of the Fisher matrix. Inversion of an ill-conditioned matrix can result in large errors on the covariance coefficients if the elements of the Fisher matrix have been estimated with insufficient precision. The conditioning number is a metric providing a mathematical lower limit to the required precision for a reliable inversion, but it is often too stringent in practice for Fisher matrices with size larger than $2\times2$. 
In this paper we propose a general numerical method to guarantee a certain precision on the inferred constraints, like the FoM. It consists on randomly vibrating (perturbing) the Fisher matrix elements with Gaussian perturbations of a given amplitude, and then evaluating the maximum amplitude that keeps the FoM within the chosen precision.
The steps used in the numerical derivatives and integrals involved in the calculation of the Fisher matrix elements can then be chosen accordingly in order to keep the precision of the Fisher matrix elements below this maximum amplitude. We illustrate our approach by forecasting stage IV spectroscopic surveys cosmological constraints from the galaxy power spectrum. We infer the range of steps for which the Fisher matrix approach is numerically reliable.
We explicitly check that using steps larger by a factor two produce an inaccurate estimation of the constraints.
We further  validate our approach by comparing the Fisher matrix contours to those obtained with an MCMC approach 
-- in the case where the MCMC posterior distribution is close to a Gaussian
-- finding excellent agreement between the two approaches.\\}

\maketitle

\section{Introduction}

Since the discovery of the acceleration of the expansion of the Universe in the late $90s$ \citep{riess_observational_1998,perlmutter_measurements_1999}, the $\Lambda$CDM model remains the most successful model in cosmology. It provides a simple and accurate description of the properties of the Universe, with a very limited number of parameters that are now constrained at the few percent level using measurements from the \emph{Planck} CMB mission \citep{Aghanim:2018eyx} and other experiments (see e.g. \citet{Percival:2001hw, 2011MNRAS.418.1707B, 2017MNRAS.470.2617A, 2018PhRvD..98d3526A}).
The accelerated expansion of the Universe can be explained by postulating a new form of matter or mechanism dubbed ``dark energy'', this component coming in addition to the dark matter. Dark energy 
dominates the energy budget of the Universe today, making up for $\sim 70 \%$  of its total energy density.

During the past two decades a variety of dark energy experiments have been proposed in order to study the observed cosmic acceleration of the Universe through various probes: optical spectroscopic and photometric galaxy clustering, weak gravitational lensing, supernovae, galaxy clusters being the most commonly used. In order to quantify the performance of a survey and in particular its ability to constrain the properties of dark energy, the Dark Energy Task Force (DETF) defined a metric, the figure of merit  \citep[FoM,][]{albrecht_report_2006}: a figure inversely proportional to the surface  bounded by the confidence contours  for the $w_0$ and $w_a$ parameters from the Chevalier-Polarski-Linder (CPL) parameterisation \citep{chevallier_accelerating_2001,linder_exploring_2003}. The FoM is now standardly  used to  quantify the performance of a particular experiment in constraining the dark energy parameters, 

We are currently entering the era of high-precision cosmology with a suite of stage IV experiments about to be deployed.  
Currently ongoing and forthcoming ground based experiments as well as space missions include DESI\footnote{\url{www.desi.lbl.gov}} \citep{desi_collaboration_desi_2016}, \Euclid\ \footnote{\url{www.euclid-ec.org}}
\citep{laureijs_euclid_2011,amendola_cosmology_2016,Euclid2019}, LSST\footnote{\url{www.lsst.org}}
\citep{lsst_science_collaboration_lsst_2009,ivezic_lsst:_2019,Chisari:2018vrw}, 
WFIRST\footnote{\url{https://wfirst.gsfc.nasa.gov/}} \citep{akeson_wide_2019,Dore:2019pld}, and the SKA\footnote{\url{www.skatelescope.org/}} \citep{Bacon:2018dui}. Forecasting activities are the key for predicting future constraints on cosmological parameters and the corresponding dark energy FoM. Indeed, forecasts not only allow to predict the performance of a future survey but also help designing and optimising it.

There exist different ways to compute forecasts. For instance one can use a simulated realization of the data and use a Monte Carlo Markov Chain (MCMC) method to sample the likelihood of the parameters of interest. This approach is  known to provide  reliable estimations of the constraints on the cosmological parameters  \citep[see e.g.][]{Dunkley:2004sv}. However, it is often quite time consuming, especially if one wishes to explore a wide variety of cases.
The Fisher matrix formalism as adopted by the DETF, is a computationally fast approach, which allows to get constraints in a few seconds or minutes. However, this method is technically touchy as the covariance matrix is obtained by inversion of the Fisher matrix, which coefficient calculations  often  require the computation of derivatives and integrals which can lead to inaccurate results, e.g. if the step sizes (as shown in later sections) are not chosen properly to ensure the appropriate precision. Moreover, the Fisher formalism can miss contour shapes and thereby potential degeneracies (see, for example \citealt{ wolz_validity_2012}) as it assumes a Gaussian likelihood in frequentist approach or  a Gaussian  posterior  of the parameters in Bayesian approach; that might not be true if, in particular, parameters are not tightly constrained by the experiment(s). Several methods have been proposed to handle these complex situations \citep{2011MNRAS.416.1010J,2014MNRAS.441.1831S,Sellentin_2015,Amendola_2016}.

In the present study we propose a numerical approach that is independent of the specific problem treated, with the purpose to test the numerical reliability of constraints obtained within the Fisher matrix formalism. As a working example, we consider the spectroscopic galaxy clustering probe for the following stage IV surveys: DESI, \Euclid\ and WFIRST-2.4. The paper is organised as follows: in Section~\ref{SEC2} we give a brief description of the aforementioned surveys. In Section~\ref{SEC3} we describe the different tests performed in order to validate the Fisher approach: namely, we determine the required precision to compute a valid Fisher matrix, carry out several stability and convergence tests, and perform a comprehensive comparison with the MCMC approach. In Section~\ref{SEC4} we present the results and address the following question: How can one numerically validate the Fisher matrix approach? Finally, we report our conclusions in Section~\ref{SEC5}.

\section{Surveys}
\label{SEC2}

\subsection{\Euclid}

\Euclid\ is a medium-class ESA mission with the main objective to understand the Universe's expansion, dark energy, and dark matter. \Euclid\ will measure the shape of more than one billion galaxies and provide tens of millions of spectroscopic redshifts. \Euclid\ will use two main cosmological probes, namely galaxy clustering and weak gravitational lensing. These will allow to probe the dark sector of the Universe with unprecedented precision. 
 \Euclid\ \citep{laureijs_euclid_2011} is targeting a FoM of 400 and is also aiming to distinguish between different theories of gravity, i.e. testing general relativity against alternative models\textbf{,}
by measuring the exponent of the growth factor $\gamma$ with an precision better than $0.02$ at $1\sigma$. \Euclid\ will also measure the neutrino masses with a $1\sigma$ precision better than $0.03$ eV. Combined with Planck, \Euclid\ will also probe some inflation models through the measurement of the non-Gaussianity of initial conditions. Two surveys are planned: a wide one of $15,000 \, {\rm deg}^2$ and a deep one of $40 \, {\rm deg}^2$. The $1.2$m telescope contains two instruments that will allow to observe around 2 billion galaxies and produce 50 million precise redshift measurements. \Euclid's large format visible imager (VIS), will probe the photometric galaxy clustering in the optical wavelength range. The near-infrared imaging and spectroscopy (NISP) instrument will perform photometry imaging and produce spectra in the near infrared range. 

\subsection{DESI (Dark Energy Spectroscopic Instrument)}

Known as the successor of the stage III BOSS and eBOSS surveys \citep{dawson_baryon_2013, Dawson_2016}, DESI is a Stage IV ground-based dark energy experiment whose primary objectives are measuring the baryonic acoustic oscillations (BAO) and constraining the growth of structure through redshift space distortions (RSD) measurements. DESI will allow to estimate the deviations from Gaussianity through the $f_{NL}$ parameter. DESI will also measure the neutrinos masses with a precision of 0.02 eV at $1\sigma$, for a maximum scale $k_{max} < 0.2$ h.Mpc$^{-1}$. DESI will scan a $14,000 \, {\rm deg}^2$ sky area and measure more than 30 million spectra from $4$ galaxy populations: the bright galaxies (BGS) at low redshift (until $z \leq 0.6$), the luminous red galaxies (LRGs) composed of highly biased objects at intermediate redshift $z < 1$, the bright emission lines galaxies (ELGs) up to $z = 1.7$ (which can only be detected at high resolution since the OII doublet must be well resolved), and the quasi-stellar objects (QSOs). The latter allow tracing both the matter distribution at high redshift and the neutral hydrogen by the $Ly-\alpha$ absorption in the spectra. DESI is now underway and will observe during five years (2019-2023) using the Mayall $4$-meter telescope at Kitt Peak.

\subsection{WFIRST (Wide Field Infrared Survey Telescope)}

WFIRST is a NASA near-infrared imaging and low-resolution spectroscopy observatory that, similarly to \Euclid, aims to address fundamental questions about the accelerated expansion of the Universe. WFIRST will determine the expansion history of the Universe and structure growth in order to constrain dark energy and modified gravity models. To estimate the dark energy equation of state, WFIRST will perform weak lensing, supernovae distances and baryonic acoustic oscillations measurements.
In this work we use the spectroscopic probe of WFIRST-2.4, a $2,000 \, {\rm deg}^2$ survey that will observe more than 20 million galaxies in the redshift range $1 \leq z \leq 3$. 

\section{Methodology}
\label{SEC3}

\subsection{The surveys combination specification}

In general, using a single probe of a stage IV survey does not constrain cosmological parameters well enough to get a Gaussian posterior distribution, in which case testing simply the Fisher formalism is meaningless. However, when the constraints are very tight, the posterior is likely to be close to gaussian. For this reason we have chosen to investigate the Fisher formalism in a case were one can anticipate gaussian posterior (an assumption that is checked by the MCMC forecast used for the full validation of the method). In order to substantially improve the constraints, we will therefore combine \Euclid, DESI and WFIRST-2.4. 
However we note that DESI and WFIRST-2.4 cover \Euclid's entire redshift range. Hence, if one wants to fully and consistently combine these three surveys, computing the cross-correlation power spectra between those surveys is necessary. For the purposes of our study we do not wish to consider the cross spectra, therefore we have to remove all the DESI and WFIRST-2.4 redshift bins that overlap with \Euclid. In practice, we  perform the full spectroscopic \Euclid\ forecasts in the redshift range of $0.9 \leq z \leq 1.8$, the DESI forecasts at low redshift $0 < z < 0.9$ and the WFIRST-2.4 forecasts at high redshift $1.8 < z < 2.7$. Moreover, between $z=0.6$ and $z=0.9$ the ELGs, LRGs and QSOs populations from DESI also overlap. Each of these populations has a different bias, and  their cross power spectra would have to be computed as well. For simplicity we only consider the ELGs population at $z=\{0.6,0.9\}$ because it has the highest density of galaxies and provides potentially the best constraints over the parameters. At $z<0.6$ we consider the BGS population.  

We consider $4$ redshift bins for the DESI survey with the following binning: $2$ redshift bins for the BGS population centered at $z =\{0.125, 0.375\}$ with a redshift size $\Delta z = 0.25$ and 2 redshift bins for the ELGs population centered at $z =\{0.6, 0.8\}$ with $\Delta z = 0.2$. Both \Euclid\ and WFIRST-2.4 are utilising $3$ redshift bins with $\Delta z = 0.3$, centered at $z = \{1.05, 1.35, 1.65\}$ and $z = \{1.95, 2.25, 2.55\}$, respectively. The full binning procedure is summarized in Table \ref{Red_R1}.

\begin{table}
\centering
\begin{tabular}{c|cccc}
  \hline
  Survey & $z_{\rm min}$ & $z_{\rm mean}$ & $z_{\rm max}$ & $dN(z_{mean})/d\Omega dz[deg^{-2}]$ \\
  \hline \hline
  \multirow{2}{*}{DESI(BGS)} & 0 & 0.125 & 0.25 & 3029 \\
  \cline{2-5}
  &0.25 & 0.375 & 0.5 & 528 \\ 
  \hline
  \multirow{2}{*}{DESI(ELGs)} & 0.5 & 0.6 & 0.7 & 309 \\
  \cline{2-5}
  &0.7 & 0.8 & 0.9 & 2124 \\ 
  \hline \hline
  \multirow{3}{*}{\Euclid} & 0.9 & 1.05 & 1.2 & 2287 \\ 
  \cline{2-5}
  & 1.2 & 1.35 & 1.5 & 1574 \\ 
  \cline{2-5}
  & 1.5 & 1.65 & 1.8 & 764 \\ 
  \hline \hline
  \multirow{3}{*}{WFIRST-2.4} & 1.8 & 1.95 & 2.1 & 6718 \\
  \cline{2-5}
  & 2.1 & 2.25 & 2.4 & 1368 \\ 
  \cline{2-5}
  & 2.4 & 2.55 & 2.7 & 781 \\ 
  \hline \hline
\end{tabular}
\caption{Redshift range, binning choices, and density of galaxes for the surveys used in this work. For more details on survey specifications see Sec~\ref{SEC3}.}
\label{Red_R1}
\end{table}

\subsection{Two approaches towards Fisher matrix cosmological constraints}

In this work, we will follow two approaches for the Fisher matrix forecast of cosmological constraints using the galaxy clustering probe. We describe these two approaches, dubbed M1 and M2, below. 

\subsubsection{Approach M1}

In the first approach, M1, the cosmological model  considered is an extension of the $\Lambda$CDM model in which we assume a variable dark energy equation of state parameter described by the CPL parameterisation. 
Furthermore, we allow for non-flatness by letting $\Omega_{\rm DE}$ vary. We also assume massless neutrinos. The full set of cosmological parameters is: \\

\noindent $\theta_{\rm cosmo} : \{\Omega_{\rm b}, h, \Omega_{\rm m}, n_{\rm s}, \Omega_{\rm DE}, w_0, w_a, \sigma_8\}$\,. \\

\noindent The cosmological parameters are the quantities of direct interest here. $\Omega_{\rm b}, h, \Omega_{\rm m}, n_{\rm s}$ are parameters that modulate the \emph{shape} of the power spectrum. $\Omega_{\rm b}$ is the baryon density parameter. $h$ is the reduced Hubble constant defined as $h = H_{\rm 0}/(100$km.s$^{-1}$.Mpc$^{-1})$ where $H_{\rm 0}$ is the present day Hubble parameter. $\Omega_{\rm m}$ is the matter density parameter, and $n_{\rm s}$ the spectral scalar index. $\Omega_{\rm DE}$ is the dark energy content directly linked to the curvature of the universe, and ($w_0$,$w_a$) represent respectively the dark energy equation of state at $z=0$ and its derivative with respect to scale factor at $z=0$, in the so-called CPL form \citep{chevallier_accelerating_2001,linder_exploring_2003}. The equation of state being then given by :
\begin{equation}
w(z) = w_0 + w_a\frac{z}{1+z} \, .
\label{Eq1}
\end{equation}

\noindent The $\sigma_8$ parameter gives the amplitude of matter fluctuations in the linear regime on a scale of  $8h^{-1}$Mpc. Furthermore, at each redshift bin we consider two \emph{nuisance} parameters: the logarithm of the product between the bias and $\sigma_8(z)$, i.e.  $\ln(b\sigma_8$), and the (residual) shot noise $P_{\rm shot}(z)$ due to the finite self-pair counts in two-point statistics. The shot noise affects the power spectrum by adding a white systematic component increasing thus the statistical uncertainties. We consider $10$ redshift bins $z_i$, thus the total number of nuisance parameters amounts to $20$: \\

\noindent $\theta_{\rm nuisance} = \{\ln(b\sigma_8(z_i)), P_{\rm shot}(z_i)\}$ \, . \\

The cosmological parameters fiducial values are summarized in Table~\ref{Red_R2}, and the nuisance parameters are given in Table~\ref{Red_R3}.

\begin{table}
\centering
\begin{tabular}{cccccccc} 
\hline
  \multicolumn{8}{c}{Cosmological parameters} \\
  \hline
  $\Omega_{\rm b}$ & $h$ & $\Omega_{\rm m}$ & $n_{\rm s}$ & $\Omega_{\rm DE}$ & $w_0$ & $w_a$ & $\sigma_8$ \\
  \hline
  0.05 & 0.67 & 0.32 & 0.96 & 0.68 & -1 & 0 & 0.83 \\ 
  \hline \hline \end{tabular}
\caption{Fiducial values for the cosmological parameters (M1 approach).}
\label{Red_R2}
\end{table}

\begin{table}
\centering
\setlength\tabcolsep{1pt}
\begin{tabular}{|c|c|} \hline
\multicolumn{2}{|c|}{Nuisance parameters} \\
  \hline
  $\ln(b\sigma_8(0.125)$ & -0.157865 \\
  \hline
  $\ln(b\sigma_8(0.375)$ & 0.052047 \\
  \hline
  $\ln(b\sigma_8(0.6)$ & -0.830210 \\
  \hline
  $\ln(b\sigma_8(0.8)$ & -0.947904 \\
  \hline
  $\ln(b\sigma_8(1.05)$ & -0.305297 \\
  \hline
  $\ln(b\sigma_8(1.35)$ & -0.293715 \\
  \hline
  $\ln(b\sigma_8(1.65)$ & -0.302905 \\
  \hline
  $\ln(b\sigma_8(1.95)$ & -0.321325 \\
  \hline
  $\ln(b\sigma_8(2.25)$ & -0.342024 \\
  \hline
  $\ln(b\sigma_8(2.55)$ & -0.364400 \\
  \hline
  $P_{\rm shot}(z)$ & 0.0 \\
  \hline \end{tabular}
\caption{Fiducial values for the nuisance parameters. They are strictly identical for both approaches: M1 and M2. The bias values are taken from \citet{desi_collaboration_desi_2016,Euclid2019,green_wide-field_2011,spergel_wide-field_2013}. The shot noise fiducial value is the same at each redshift bins.}
\label{Red_R3}
\end{table}

\subsubsection{Approach M2}

In the second approach, M2, adopted in \citet{Euclid2019}, 4 quantities which determine the shape of the power spectrum are treated as independent of the cosmological parameters. We call these 4 quantities the shape parameters. The other quantities, the angular diameter distance $D_a(z)$, the Hubble rate $H(z)$, and the growth rate of cosmic structure $f(z)$ are called the redshift dependent (rd) parameters. With the addition of $3$ redshift dependent parameters and $2$ redshift dependent nuisance parameters (the same as in M1) per redshift bin, for ten redshift bins, this amounts to $30$ redshift dependent parameters and $20$ nuisance parameters. The full set consists of $54$ parameters (see Table~\ref{Red_R4}) : \\

\noindent $\theta_{\rm shape} : \{\omega_{\rm b}, h, \omega_{\rm m}, n_{\rm s}\}$ \\

\noindent $\theta_{\rm rd+nuisance}: \{\ln(D_a(z_i)), \ln(H(z_i)), \ln(f\sigma_8(z_i)), \ln(b\sigma_8(z_i)), P_{\rm shot}(z_i)\}$. \\  

\noindent 
The nuisance parameters are identical in the two approaches. As in \citet{Euclid2019}, we estimate the constraints of the physical baryon density ($\omega_{\rm b}=\Omega_{\rm b}h^2$), and the physical matter density ($\omega_{\rm m}=\Omega_{\rm m}h^2$) parameters. In the M2 approach, the constraints on the final cosmological parameters ($\theta_{\rm cosmo}$) are obtained from the $54$ parameters stated above by projection.
We also note that taking the logarithm of the parameters helps to stabilise matrix inversion by reducing the dynamic range between minimum and maximum eigenvalues, as described in \citet{Euclid2019}.

\begin{table}
\centering
\setlength\tabcolsep{1pt}
\begin{tabular}{|c|c||c|c|}
  \hline
  \multicolumn{4}{|c|}{Cosmological and \textbf{redshift dependent parameters}} \\
  \hline
  $\omega_{\rm b}$ & 0.022445 & $h$ & 0.67 \\
  \hline
  $\omega_{\rm m}$ & 0.143648 & $n_{\rm s}$ & 0.96 \\
  \hline
  $\ln(D_a(0.125))$ & 6.177874 & $\ln(H(0.125))$ & 4.268284 \\
  \hline
  $\ln(f\sigma_8(0.125))$ & -0.757651 & $\ln(D_a(0.375))$ & 7.00907 \\
  \hline
  $\ln(H(0.375))$ & 4.411368 & $\ln(f\sigma_8(0.375))$ & -0.714737 \\
  \hline
  $\ln(D_a(0.6))$ & 7.264910 & $\ln(H(0.6))$ & 4.548941 \\
  \hline
  $\ln(f\sigma_8(0.6))$ & -0.730402 &  $\ln(D_a(0.8))$ & 7.378986 \\
  \hline
  $\ln(H(0.8))$ & 4.672001 & $\ln(f\sigma_8(0.8))$ & -0.767895 \\
  \hline
  $\ln(D_a(1.05))$ & 7.452405 & $\ln(H(1.05))$ & 4.821969 \\
  \hline
  $\ln(f\sigma_8(1.05))$ & -0.830894 & $\ln(D_a(1.35))$ & 7.488258 \\
  \hline
  $\ln(H(1.35))$ & 4.992418 & $\ln(f\sigma_8(1.35))$ & -0.916615 \\
  \hline
  $\ln(D_a(1.65))$ & 7.494160 & $\ln(H(1.65))$ & 5.150878 \\
  \hline
  $\ln(f\sigma_8(1.65))$ & -1.004727 & $\ln(D_a(1.95))$ & 7.483737 \\
  \hline
  $\ln(H(1.95))$ & 5.297446 & $\ln(f\sigma_8(1.95))$ & -1.090957 \\
  \hline
  $\ln(D_a(2.25))$ & 7.463977 & $\ln(H(2.25))$ & 5.432989 \\
  \hline
  $\ln(f\sigma_8(2.25))$ & -1.173509 & $\ln(D_a(2.55))$ & 7.438762 \\
  \hline
  $\ln(H(2.55))$ & 5.558599 & $\ln(f\sigma_8(2.55))$ & -1.251760 \\
  \hline
\end{tabular}
\caption{Fiducial values of the shape and redshift dependent parameters for the approach M2.}
\label{Red_R4}
\end{table}

\subsection{The Fisher matrix formalism}

The Fisher matrix formalism \citep[see][for a pedagogical introduction]{coe_fisher_2009} is commonly  used in cosmology to forecast the Gaussian uncertainties of a set of model parameters under some constraints. The Fisher approach relies on the assumption of Gaussian posterior distribution. The Fisher matrix corresponding to the information on cosmological parameters provided by a galaxy clustering survey is given by \citep{tegmark_measuring_1997,tegmark_measuring_1998}: 
\begin{multline}
F_{\alpha\beta} = \frac{1}{8\pi^2} \int_{-1}^{1} d\mu \int_{k_{\rm min}}^{k_{\rm max}} k^2dk \\ \times \frac{\partial \ln P_{\rm obs}(z,k,\mu)}{\partial \alpha} \frac{\partial \ln P_{\rm obs}(z,k,\mu)}{\partial \beta}V_{\rm eff}(z,k,\mu) \, ,
\label{Eq2}
\end{multline}

\noindent 
where $\alpha$ and $\beta$ run over the parameters we vary. $k$ is the total wave vector magnitude in Mpc$^{-1}$, $\mu$ the cosine of the angle to the line-of-sight, and $V_{\rm eff}(z,k,\mu)$ the effective volume of the survey:

\begin{equation}
V_{\rm eff}(z,k,\mu) = V_{\rm s}(z)\left[\frac{n(z)P_{\rm obs}(z,k,\mu)}{n(z)P_{\rm obs}(z,k,\mu)+1}\right]^2 \, ,
\label{Eq3}
\end{equation}

\noindent with $n(z)$ the number density of galaxies in each redshift bin and $V_{\rm s}(z)$ the volume of each redshift bin. The  observed power spectrum $P_{\rm obs}(z,k,\mu)$, taking into account nonlinear effects is given by:

\begin{multline}
 P_{\rm obs}(z,k,\mu) = \frac{1}{q_{\perp}^2q_{\parallel}} \left(\frac{[b\sigma_8(z)+f\sigma_8(z)\mu^2]^2}{1 + [f(z)k\mu \sigma_p(z)]^2}\right) \\ \times \frac{P_{\rm dw}(z,k,\mu)}{\sigma_8^2(z)}F_z(z,k,\mu) + P_{\rm shot}(z) \, .
 \label{Eq4}
\end{multline}

\noindent The first term represents the Alcock-Paczynski (AP) volume dilation effect that reflects spurious anisotropies in the power spectrum  which arises when  assuming a cosmology (that might be different from the true cosmology) to convert redshifts into distances. The coefficient $q_\parallel=H_{\rm ref}(z)/H(z)$ is given by the ratio between the Hubble parameter in the true cosmology and the one in the reference cosmology. The coefficient $q_\perp=D_a(z)/D_{\rm a,ref}(z)$ is given by the ratio between the angular distance in the reference cosmology and the one in the true cosmology. The term, between parentheses, contains the linear redshift space distortions (RSD) due to the growth of structure, and finger-of-god effects that aim to describe the nonlinear damping due to the velocity dispersion of satellite galaxies inside host halos. $\sigma_{\rm p}=1/(6\pi^2)\int P_m(k,z)dk$ represents the linear-theory velocity dispersion that can be computed from the linear matter spectrum $P_{\rm m}(k,z)$. The third term, $P_{\rm dw}$, is the dewiggled power spectrum \citep{eisenstein_baryonic_1998,eisenstein_power_1999} that takes into account nonlinearities in the matter power spectrum. The fourth term, $F_z$, represents the uncertainties coming from the spectroscopic precision of the survey under consideration and $P_{\rm shot}(z)$ the residual shot noise. For a comprehensive description of the observed power spectrum model introduced above we recommend the recent \Euclid\ paper on validated forecasts \citep{Euclid2019}, where the specific form of the function $F_z$ as well as the rescaling of $k$ and $\mu$ due to the AP effect are given in  Equations~(74)-(79). In terms of nonlinearities, we should note that the model used accounts for the finger-of-God effect as well as for the damping of the BAO feature. More details can be found in Section 3.2.2 of \citet{Euclid2019}.

We move on to describe the estimation of the uncertainties using the Fisher matrix formalism.
The inverse of the Fisher matrix  gives the covariance matrix that represents the likelihood curvature evaluated at the fiducial values of $\alpha$ and $\beta$. According to the Cram\'er-Rao inequality, the square root of each covariance matrix diagonal elements gives the $1\sigma$ lower bound constraint for each parameter (marginalised over all other parameters) where the posterior distribution is Gaussian:

\begin{equation}
\sigma_{\alpha} = \sqrt{(F^{-1})_{\alpha\alpha}} \, .
\label{Eq5}
\end{equation}
In order to compute the Fisher matrix we use \texttt{SpecSAF} (Spectroscopic Super Accurate Forecast), a modified and improved version of a code first used in \cite{tutusaus_dark_2016}. \texttt{SpecSAF} computes high-point stencil derivatives with a very high precision level. This code is linked to CAMB \citep{lewis_efficient_2000} and CLASS \citep{lesgourgues_cosmic_2011} to compute matter power spectra, and allows the user to directly compute the FoM and plot the contours. We note that \texttt{SpecSAF} is one of the validated codes used in \cite{Euclid2019}.

\subsection{Precision of the Fisher matrix formalism}

In order to obtain reliable final constraints from the covariance matrix one has to estimate the  precision needed on the Fisher elements themselves, to achieve a given precision (for instance a 
relative precision lower than $10\%$ for the square root of elements of the diagonal of the covariance matrix). This problem is related to the conditioning of the Fisher matrix that needs to be inverted. The  condition number $C_N$ of a matrix $A$ is known to provide an upper limit to the precision $\delta x$ of the determination of the solution $x$ of the equation $Ax=b$, where $b$ \textbf{is} a vector. That is, the elements of $b$ and $A$ have to be known with a precision such that $\delta b_i/b_i \leq (1/C_N) \, \delta x/x$. That is, if the condition number is large, even a very small error in $b$ will result in a large error in $x$.
The condition number \citep{belsley_regression_2005} is given by the ratio between the largest eigenvalue to the smallest eigenvalue. This is however an upper limit that can be far too stringent in practice: as an example, in our case the condition number in approach M1 is $1.3 \times 10^5$, while in the M2 approach the condition number is  $4.4 \times 10^{11}$. This formally calls for a precision of $\sim 10^{-6}$ and $\sim 10^{-13}$ to ensure an accuracy of $10\%$ in the final constraints. However, a diagonal matrix (provided that no term in the diagonal vanishes) is well conditioned in practice, even if the condition number is very large. Therefore for a matrix with size significantly greater that $2\times2$, the condition number does not provide useful information on the conditioning of the matrix. This observation has important consequences in practice, as the elements of the Fisher matrix are generally computed from numerical derivatives whose required precision has to be determined in order to achieve reliable results on the covariance. 
To better quantify the precisoin needed for reliable constraints we introduce a simple method that consists in \emph{vibrating} (perturbing) each element of the Fisher matrix as follows:

\begin{equation}
F^{\rm vibrated}_{\alpha\beta} = F_{\alpha\beta}(1 + \epsilon N(0,1)) \, ,
\label{Eq8}
\end{equation}

\noindent with $\alpha$ and $\beta$, indices which run over all unique pair of parameters. $N(0,1)$ is a normal distribution centered at $0$ with variance $1$ and $\epsilon$ the amplitude of the perturbations. In order to keep the Fisher matrix symmetric, we only perturb the lower triangle part of the Fisher and compute the symmetric part across the diagonal. The coefficients in the Fisher matrix are computed from Boltzmann codes and are therefore subjects to correlated (numerical) noise. In our approach this correlation is lost during the Fisher matrix vibration. 
In the approach M1, we apply the perturbations for $2000$ amplitudes, $\epsilon$, regularly log-spaced in the  range  $10^{-12}$ -- $10^{-1}$. To ensure enough statistical data and reduce the Poisson noise, we produce $10,000$ vibrated Fisher matrix per epsilon values. In the M2 approach this procedure is more expensive in terms of computational time. We therefore reduced the number of  $\epsilon$ values to $600$ in the range  $10^{-6}$ -- $10^{-1}$ and we produced $2000$ vibrated Fisher matrices per $\epsilon$ value. The constraints obtained with each vibrated Fisher matrix are then compared with the original Fisher matrix by computing the relative error. For each parameter we compute the number of draws that gives an agreement level on the constraints at the chosen precision ($10\%$ in our case) or better.

The tolerated precision is the largest value of $\epsilon$ for  which $68\%$ of the constraints on each parameter remain lower than the chosen precision ($10\%$ in our case). The results are illustrated in Fig.~\ref{VBR_M1} for the cosmological parameters in the M1 approach. In the M2 approach the size of the matrix is too large so in Fig.~\ref{VBR_M2} we only show the shape parameters and the redshift dependent parameters for a single redshift bin ($z=1.35$).

\subsection{Stability and Convergence tests}

The precision and stability (convergence) of numerical derivatives can be an important issue and testing it is an essential validation step. Fortunately, testing the stability of the derivative of the observed power spectrum over any parameter is straightforward and determines the appropriate step sizes to use (and which ones to avoid). However the stability itself does not prove that a numerical derivative is correctly computed since it only shows to what extent the derivative changes as a function of step size. Ideally, a convergence test (computing the relative error between the analytical derivative and the numerical one) also has to be performed. Unfortunately, directly exploiting the convergence with a relative error test is not  possible when the derivatives cannot be computed analytically. This is the case for most of the parameters we consider in this study. For the cosmological parameters we need to use a Boltzmann code, and the semi-analytical form of the derivatives over $D_a(z)$ and $H(z)$ contain derivatives of the matter power spectrum over the scale $k$ ($dP/dk$). Here we follow an alternative approach to test derivative convergence by considering three different numerical methods to compute the derivatives: the $3$, $5$ and $7$ points centered schemes. With the 3 (respectively 5, 7) scheme being the method in which we use 3 (resp. 5, 7) perturbed points of the function f(x) to determine the numerical derivative at x, we assume that the $7$ points derivative is very near the true value when an appropriate step size is chosen. We compare the $7$ points with the $3$ and $5$ points stencil for different step sizes and look for best agreement between the two derivatives, in other words we make convergence tests between the 3 and the 5 points stencil towards the 7 points stencil by comparing the relative error between the 7 points stencil and the 3 and 5 points stencils. More generally the convergence is always faster with the high stencils derivatives method than with the low stencils derivatives, which means that the optimum for the $5$ points stencil often occurs at larger step values than the $3$ points stencil. However this does not hold in all cases. For instance applying this method to a highly noisy/oscillatory function can lead to wrong results and finding stable behaviours is not possible. It is therefore very useful
to plot the functions under consideration in order to identify any such features.   

\subsection{The MCMC sampling}

In order to consolidate the reliability of our Fisher matrix constraints, we compare them with MCMC constraints. 

As is well known, the MCMC sampling is much more time-consuming than the Fisher matrix computation, 
due to the computational time needed to obtained the power spectrum at each iteration. Moreover, by construction a Markov chain cannot be parallelized, as each draw depends on the previous displacement (note, however, that it is possible to launch several chains at the same time to sample the parameter space more effectively). Considering a set of data and a model equipped with a set of parameters $\Theta$, the MCMC allows to use Bayesian inference and model comparison:

\begin{equation}
P(\Theta | {\rm data}) = \frac{P({\rm data} | \Theta) P(\Theta)}{P({\rm data})} \, ,
\label{Eq6}
\end{equation}

\noindent with $P(\Theta | {\rm data})$ the posterior distribution which provides the constraints and the joint covariance between all the parameters considered, $P({\rm data} | \Theta)$ the Likelihood, $P(\Theta)$ the prior that we take to be  flat in the present study, and $P({\rm data})$ the Bayesian evidence that can be safely ignored as it is a constant of no interest in our situation. Thus, the previous equation can be reduced to:

\begin{equation}
P(\Theta | {\rm data}) \propto P({\rm data} | \Theta) \, .
\label{Eq7}
\end{equation}

\noindent We compute the posterior distribution with the MCMC module of \texttt{SpecSAF} using the Metropolis-Hastings algorithm \citep{metropolis_monte_1949,robert_metropolis-hastings_2015}. This is described in more detail in Appendix~\ref{sec:Metropolis-Hastings}. We test the chains convergence using the Gelman-Rubin diagnostic \citep{gelman_inference_1992,brooks_general_1998} described in Appendix~\ref{sec:Gelman-Rubin}. Our data are modeled by \st{the} a realisation of our fiducial model given by the equation \ref{Eq4}. 

\subsection{MCMC Comparison}

The full validation of the Fisher matrix approach is performed with a direct comparison with the results from the MCMC chains. We sample the two parameter spaces of the two approaches and compare the relative error given by the covariance matrix of the sampling and the inverse of the Fisher matrix. We also visualize the Fisher contours versus the MCMC sampling. As we use exactly the same specifications for both methods we expect to get the same constraints with a relative error lower than $10\%$ as long as the posterior distribution remains gaussian and doesn't present any degeneracies.

\section{Results}
\label{SEC4}

\subsection{Precision needed}

\begin{figure*}[t!]
$\begin{array}{rcl}
    \includegraphics[width=0.999\textwidth]{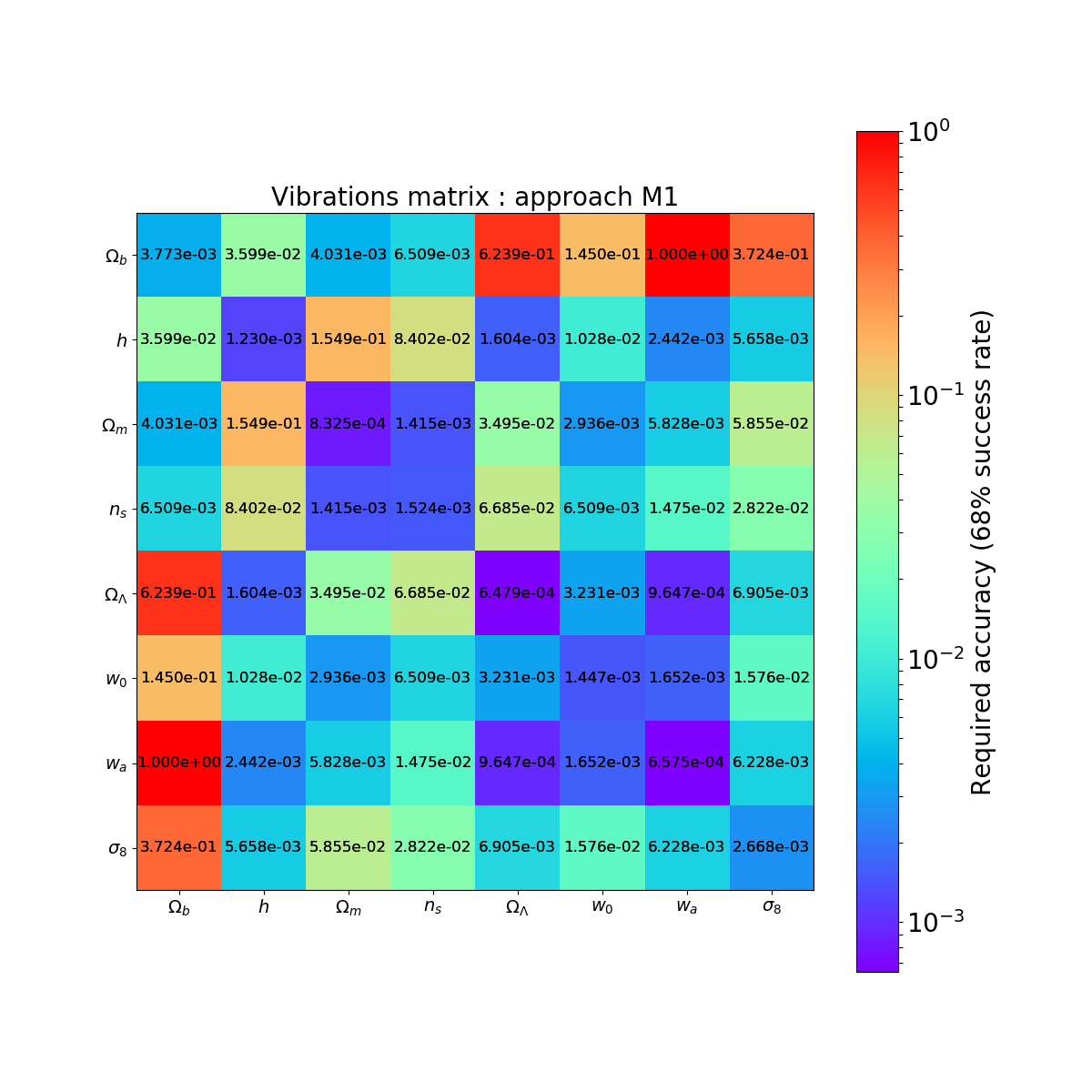} &
\end{array}$
\caption[VB_plot]{\label{VBR_M1}Vibration matrix: approach M1. The values are expressed in absolute values. Each cell containing 1 has to be taken as a lower bound value of the limit because we take 1 as a threshold perturbation amplitude.}
\end{figure*}

\begin{figure*}[t!]
$\begin{array}{rcl}
    \includegraphics[width=0.999\textwidth]{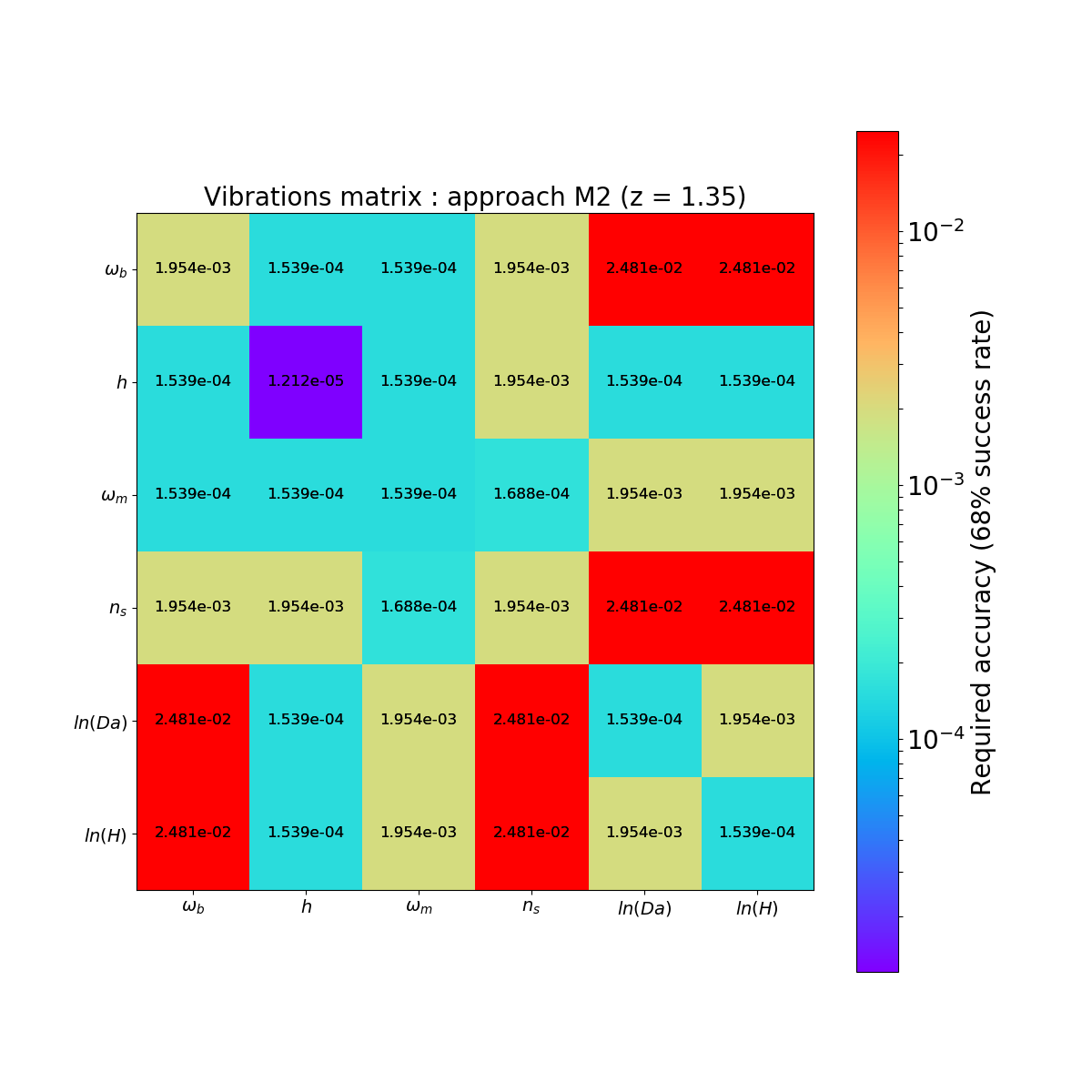} &
\end{array}$
\caption[VB_plot]{\label{VBR_M2}Vibration matrix: approach M2. Note that some values are identical because we decreased the resolution of the perturbation amplitude range.}
\end{figure*}

The vibration matrices for M1 and M2 are presented in Figure~\ref{VBR_M1} and Figure~\ref{VBR_M2}, respectively. We can clearly see that the sensitivity of the constraints depends on the element vibrated. According to Figure~\ref{VBR_M1}, the diagonal elements are the most sensitive, more specifically, $\Omega_{\rm m}$, $\Omega_{\rm DE}$ and $w_a$. For our working requested precision ($10\%$), these three parameters require a precision between $0.06\%$ and $0.09\%$; for the other parameters the diagonal elements required precision spreads between $0.1\%$ and $0.4\%$; the non-diagonal elements are less cumbersome: some of them can even change by a factor 2 without affecting the constraints (for instance $\Omega_{\rm b}$ vs [$\Omega_{\rm DE}$, $w_0$, $w_a$, $\sigma_8$]). Most of the non-diagonal elements can be computed with a precision worse than $1\%$. 

The Fisher matrix produced following the M2 approach (Figure~\ref{VBR_M2}) shows a greater sensitivity to vibration than the M1 approach, in line with their different condition numbers. The diagonal element $h$ seems to be the most sensitive parameter with a required precision around $0.0012\%$, which is however much less stringent than what the condition number would suggest ($ \sim 10^{-11}\%$). The background quantities and $\omega_{\rm m}$ require a precision around $0.015\%$, while $\omega_{\rm b}$ and $n_{\rm s}$ are one order of magnitude more tolerant with a target precision of $0.2\%$. The off-diagonal elements show a higher tolerance to perturbations, however only the background elements vs [$\Omega_{\rm b}$, $n_{\rm s}$] have a vibration tolerance higher than one percent. 

To summarise, compared to the M1 approach, the Fisher matrix produced in the M2 approach is generally at least two orders of magnitude more sensitive for the most sensitive parameters, and one order of magnitude more sensitive for the less ones. These final figures are poorly reflected by the condition\st{ing} number of each approach. 
The above vibration matrix approach allows one to quantify the precision requested for the elements in the Fisher matrix to achieve the requested precision on the constraints of interest, the FoM in our case. This effective precision is hard to anticipate otherwise: the condition number provides a mathematical lower limit to it. However, for a simple $2\times 2$ diagonal matrix, it is clear that the precision  needed for each elements is just the one wished on constraints, while the condition number (being the product of the eigenvalues) can be arbitrary large. The precision requested for the elements in the Fisher matrix is therefore intimately related to the internal structure of the Fisher matrix. \\

In the next section we examine the steps sizes necessary to achieve the requested accuracy on the Fisher matrix elements.

\subsection{Stability and convergence}

\begin{figure*}[t!]
$\begin{array}{rcl}
    \includegraphics[width=0.499\textwidth]{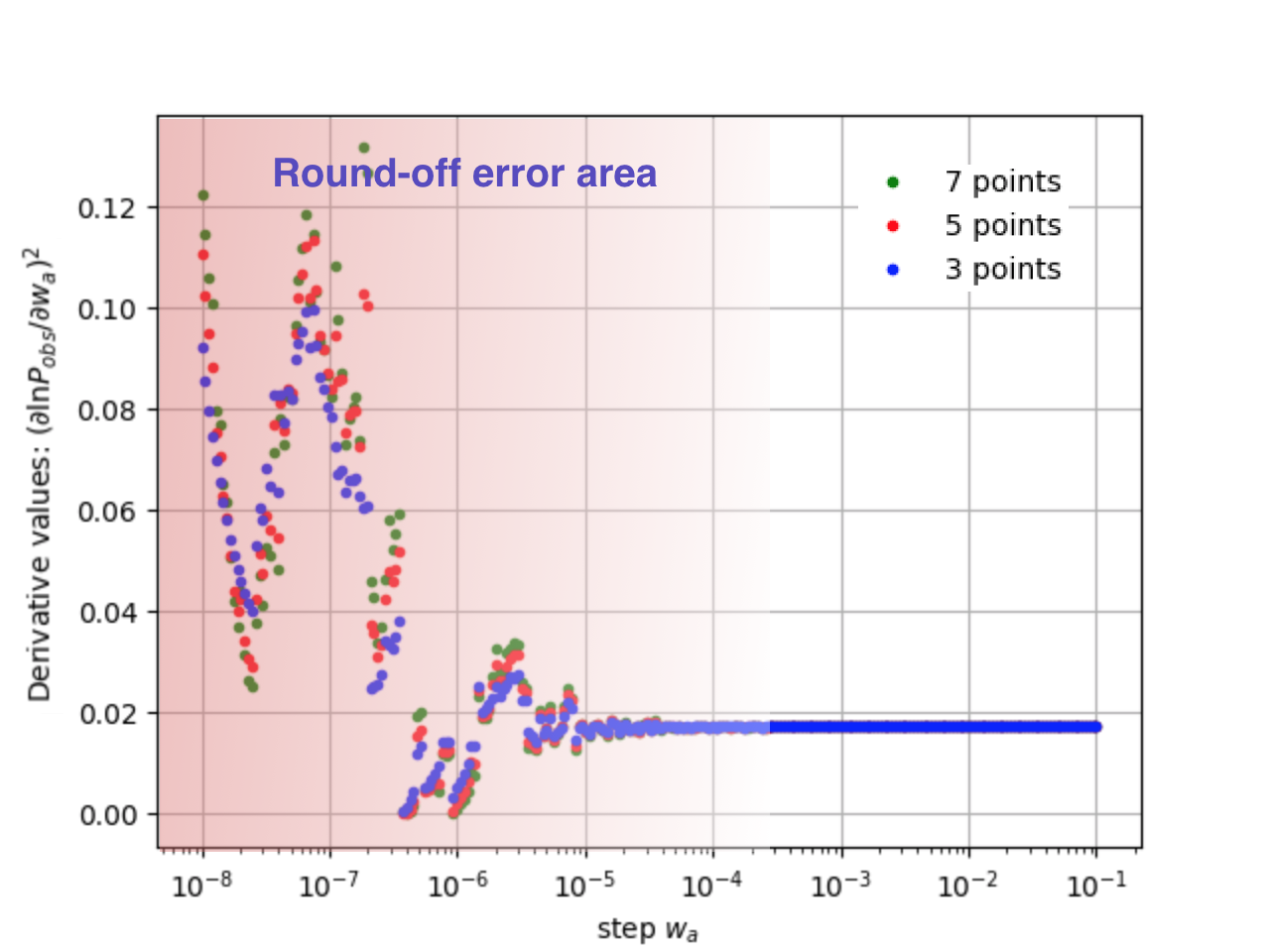} &
    \includegraphics[width=0.499\textwidth]{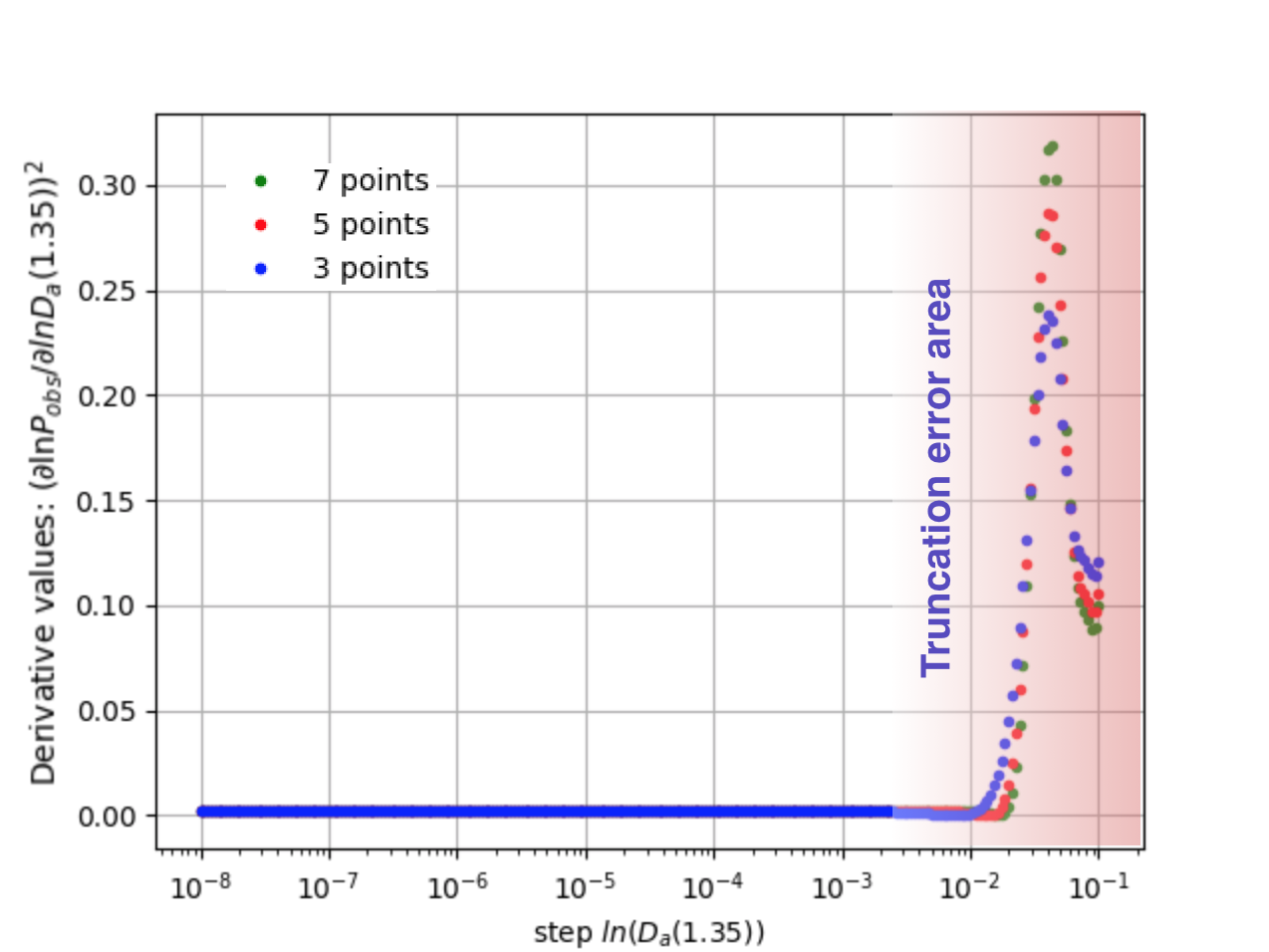} \\
    \includegraphics[width=0.499\textwidth]{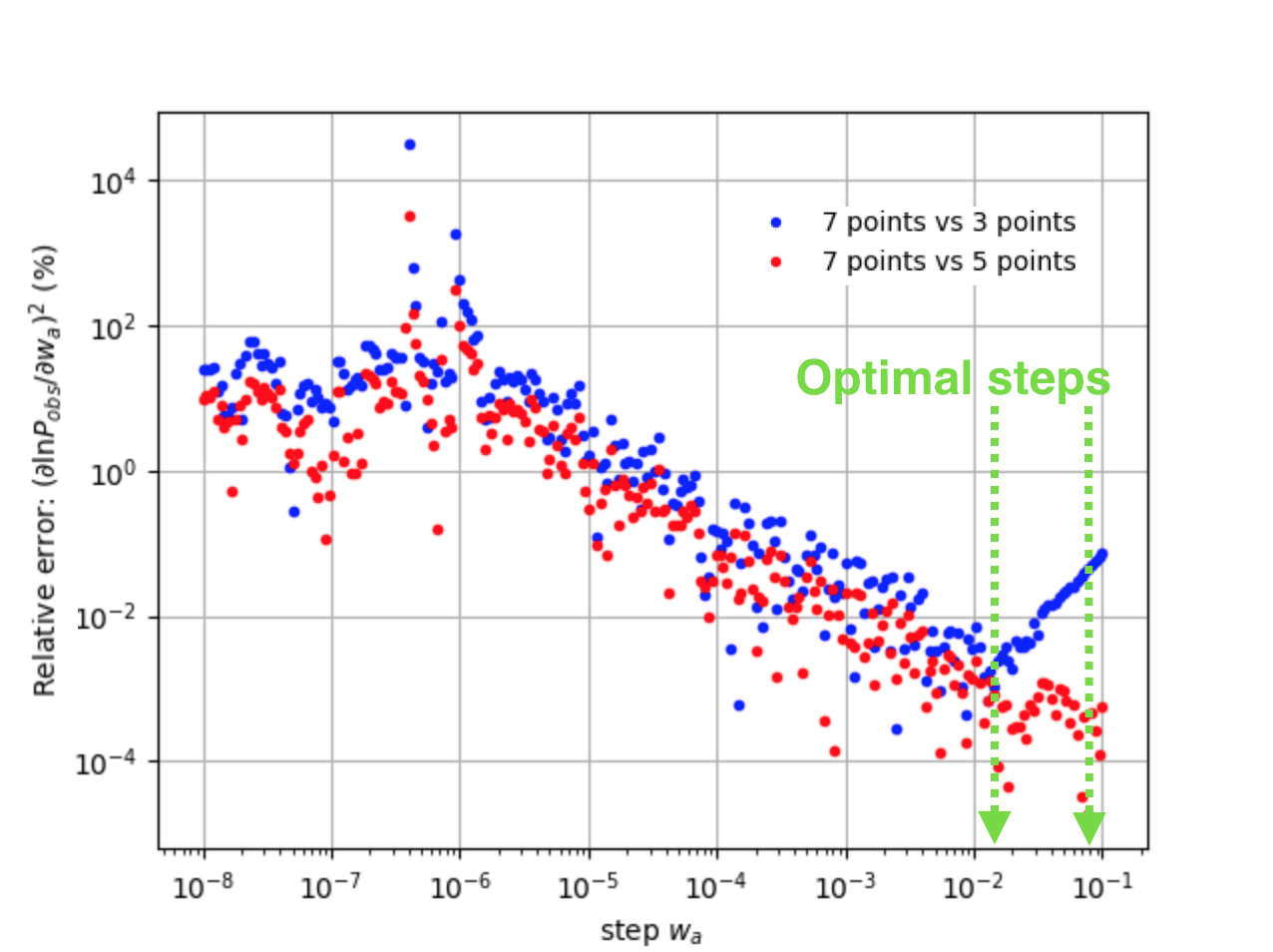} &
    \includegraphics[width=0.499\textwidth]{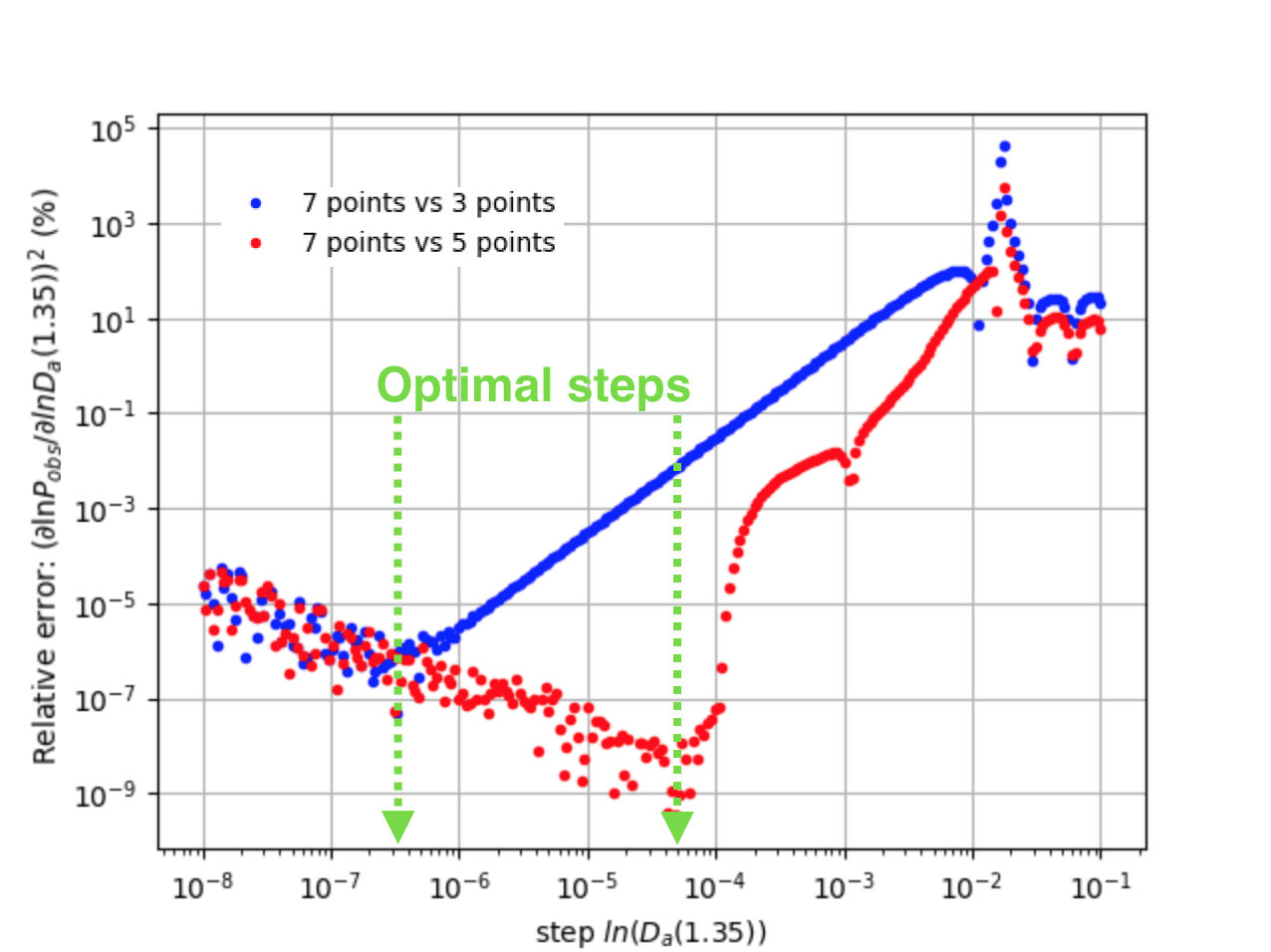} 
\end{array}$
\caption[SC_plot]{\label{SC_P} Stability and convergence tests towards the 7 points stencil for $w_a$ and ${\rm ln}(D_a(1.35))$ at a given $k$ and $\mu$ at $z=1.35$. Upper left panel: stability of $w_a$ ; upper right panel: stability of ${\rm ln}(D_a(1.35))$ ; bottom left panel: convergence of $\omega_{\rm m}$ ; bottom right panel: convergence of ${\rm ln}(D_a(1.35))$. Blue: 3 points stencil, red: 5 points stencil, green: 7 points stencil.}
\end{figure*}

In order to compute the elements of the Fisher matrix (Eq. \ref{Eq2}), one has to compute several integral quantities involving derivatives. The precision on these elements is therefore directly related to the precision one can achieve on these computations, determined by the step sizes used and the numerical schemes used in these derivations (in  the following, all the step sizes shown are given relative to the assumed fiducial value for each parameter, except $w_{\rm a}$ and $P_{shot}$ because their fiducial values are 0). Moreover, we choose to illustrate one shape parameter and one redshift dependent parameter in the following plots, because the derivatives behavior between the shapes parameters is similar as well as the derivative behavior between the redshift dependent parameters. We have therefore examined the precision obtained on each of these elements by first examining the precision reached when the step size varies. In order to illustrate this we provide illustration for derivatives against step size, which are representative of general behavior. The upper panels of Figure~\ref{SC_P} illustrate the stability of the square of the derivative of $\ln P(k,\mu)$ with respect to $w_a$ at $(k, \mu) = (0.0121, 0.5)$ and ${\rm ln}D_a$ at $(k, \mu) = (0.098, -0.1)$ (blue: 3 points stencil, red: 5 points stencil, green: 7 points stencil), at $z=1.35$. The upper left panel demonstrates the cosmological parameters stability behaviour. The stability of the derivative is reached when the slope of the derivative value over the step size is close to zero (when the curve shows a horizontal behaviour) while decreasing the step size. These derivatives are most of the time relatively stable in the step size range $[10^{-4},10^{-1}]$ for each derivative. The truncation errors remains small in all of the range for most of the derivatives. We note that for the low step values, instabilities occur due to round-off errors. The upper right panel is representative of the typical stability behaviour of ${\rm ln}(H(z))$ and ${\rm ln}(D_a(z))$. For step sizes higher than $3 \times 10^{-3}$ instabilities begin to arise for the 3 points stencil. The truncation errors dominate and increase the total error budget. The instability range for the 5 and 7 points derivatives is smaller: typically around $1 \times 10^{-2}$. The lower panels (Figure \ref{SC_P}) represent the corresponding derivative convergence tests with respect to the 7 points derivatives (relative error between the 7 points stencil and the (3, 5) points stencil in (blue, red)). Generally, convergence tests are made to compare the relative error between the analytical solution of a derivative and solutions obtained numerically when we vary the step size. The minimum relative error between the analytical form and the numerical solutions corresponds to the optimal step\,\footnote{See Appendix \ref{sec:definitions} for a summary of the main definitions relevant for this work.} ensuring the most accurate derivative. Because we cannot compute the analytical form of most of the derivatives presented here, we make convergence tests of the 5 and the 3 points stencil using the 7 points stencil results as a reference, since the latter is theoretically more accurate. In the bottom left panel, the 3 points derivative achieves its convergence when the step is equal to $10^{-2}$. The optimal step for the 5 points stencil is  around $1 \times 10^{-1}$ at most. Concerning the angular diameter distance (bottom right panel), the 3 points convergence level is achieved down to a step of $3 \times 10^{-7}$ whereas for the 5 points derivatives it is around $5 \times 10^{-5}$. The figure \ref{SC_P} shows that the convergence is always reached after the stability if we decrease the step size. For instance let's take a closer look on the figure \ref{SC_P} right panel ($\ln D_a(1.35)$). If we observe the stability and the corresponding convergence plots at a step of $10^{-1}$, we see that the convergence and the stability aren't reached yet. Now if we start to decrease the step size, we see that the stability is reached around a step of $10^{-2}$ because the slope of the curve becomes near to 0 . However, the corresponding convergence plot still shows an error of $100\%$ between the 7 points stencil and the 5 and 3 points stencils. If we continue to decrease the step size, the minimum of convergence occurs at $10^{-4}$ for the 5 points stencil and $3.10^{-7}$ for the 3 points stencil. This feature always appears as we can see it as well on the left panel of the figure \ref{SC_P} (at least for the 3 points stencil because we took a maximum step value of $10
^{-1}$). 

\begin{figure*}[t!]
$\begin{array}{rcl}
    \includegraphics[width=0.499\textwidth]{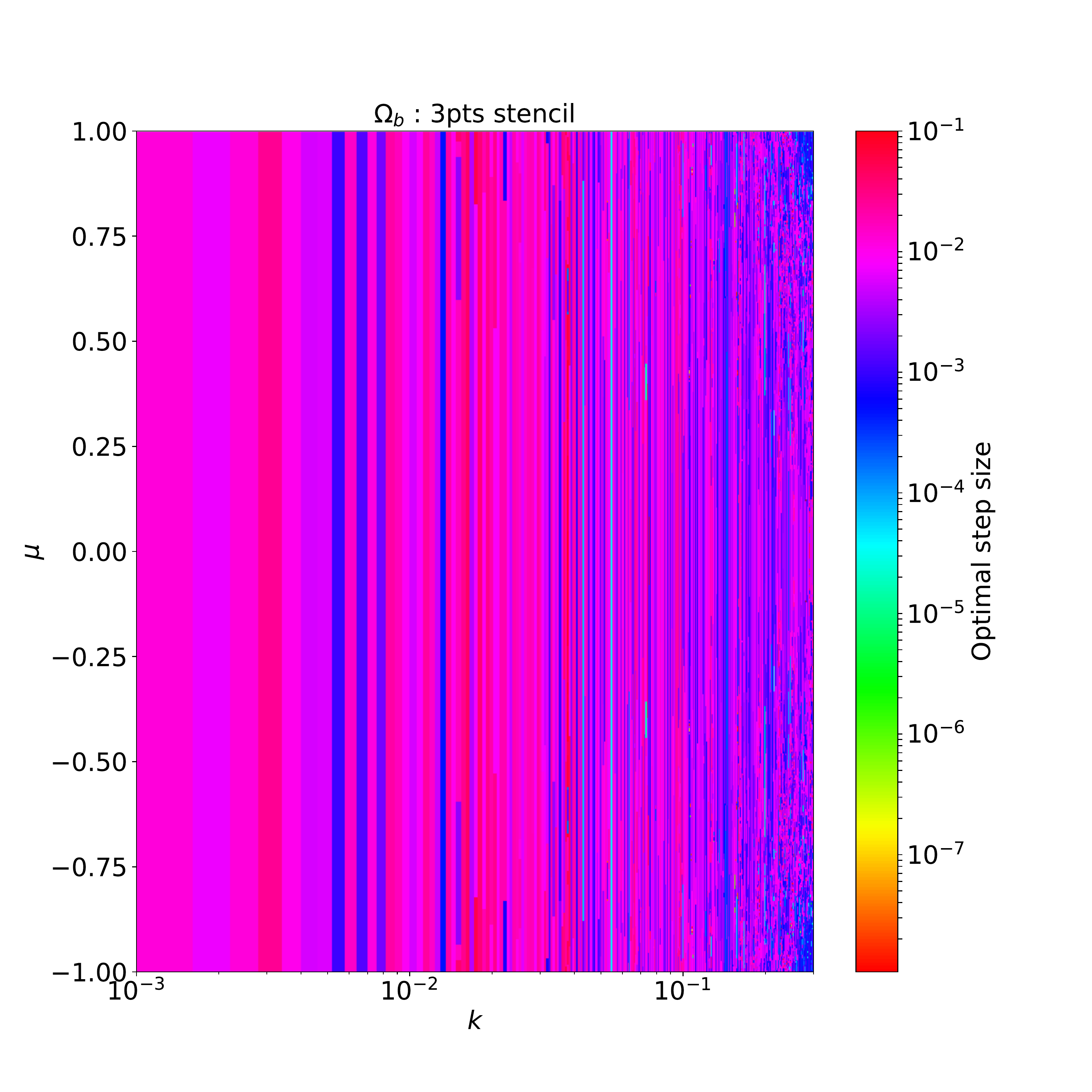} &
    \includegraphics[width=0.499\textwidth]{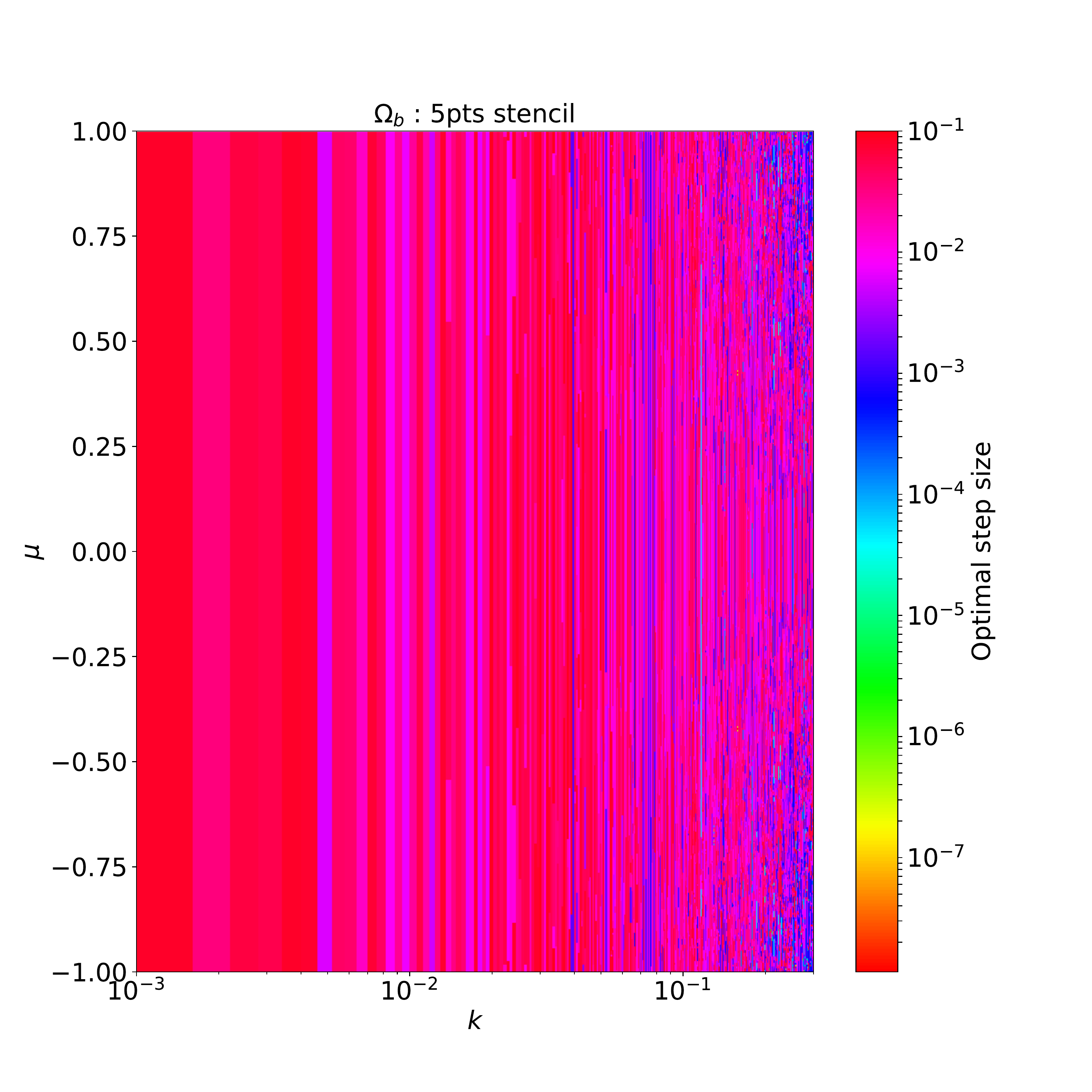} \\
    \includegraphics[width=0.499\textwidth]{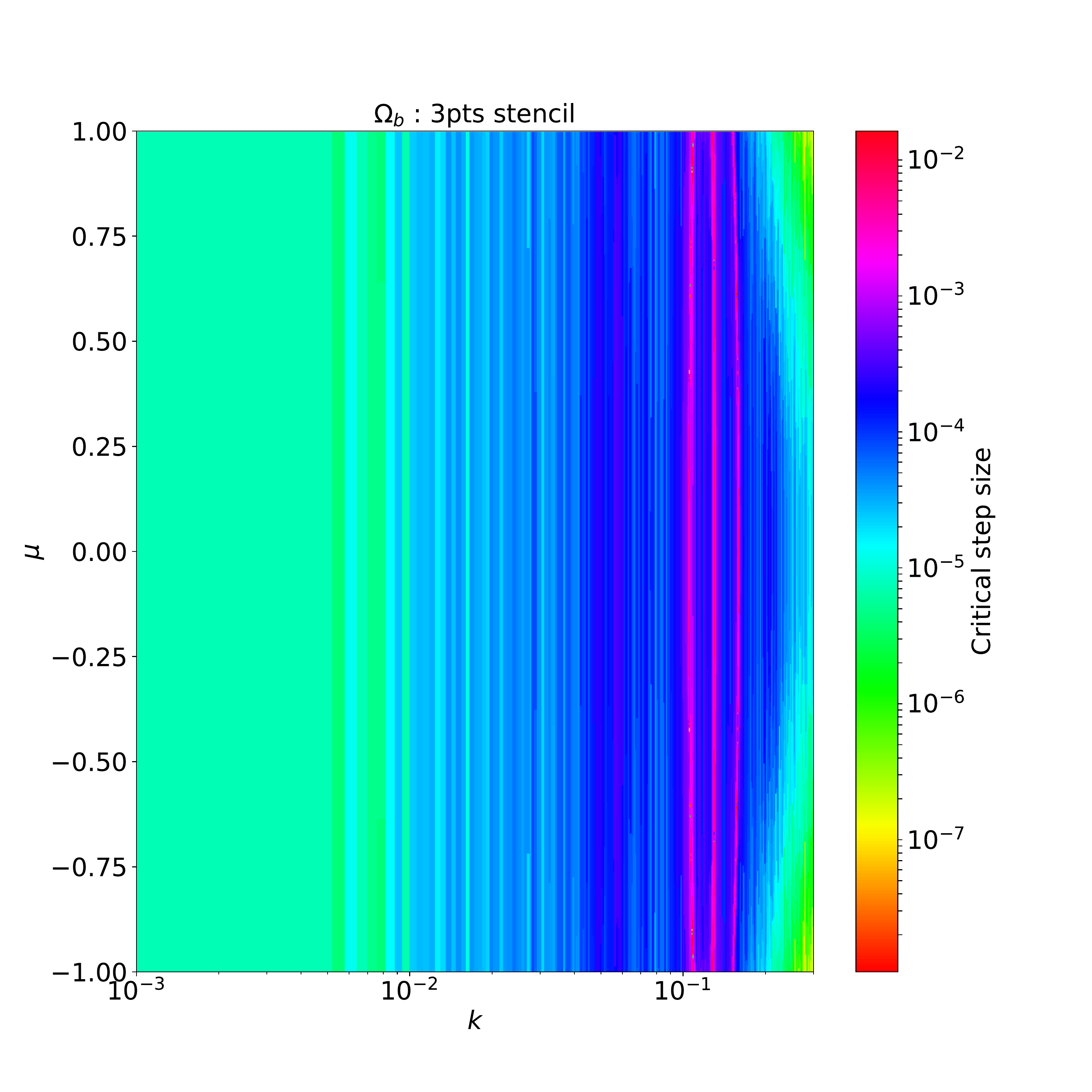} &
    \includegraphics[width=0.499\textwidth]{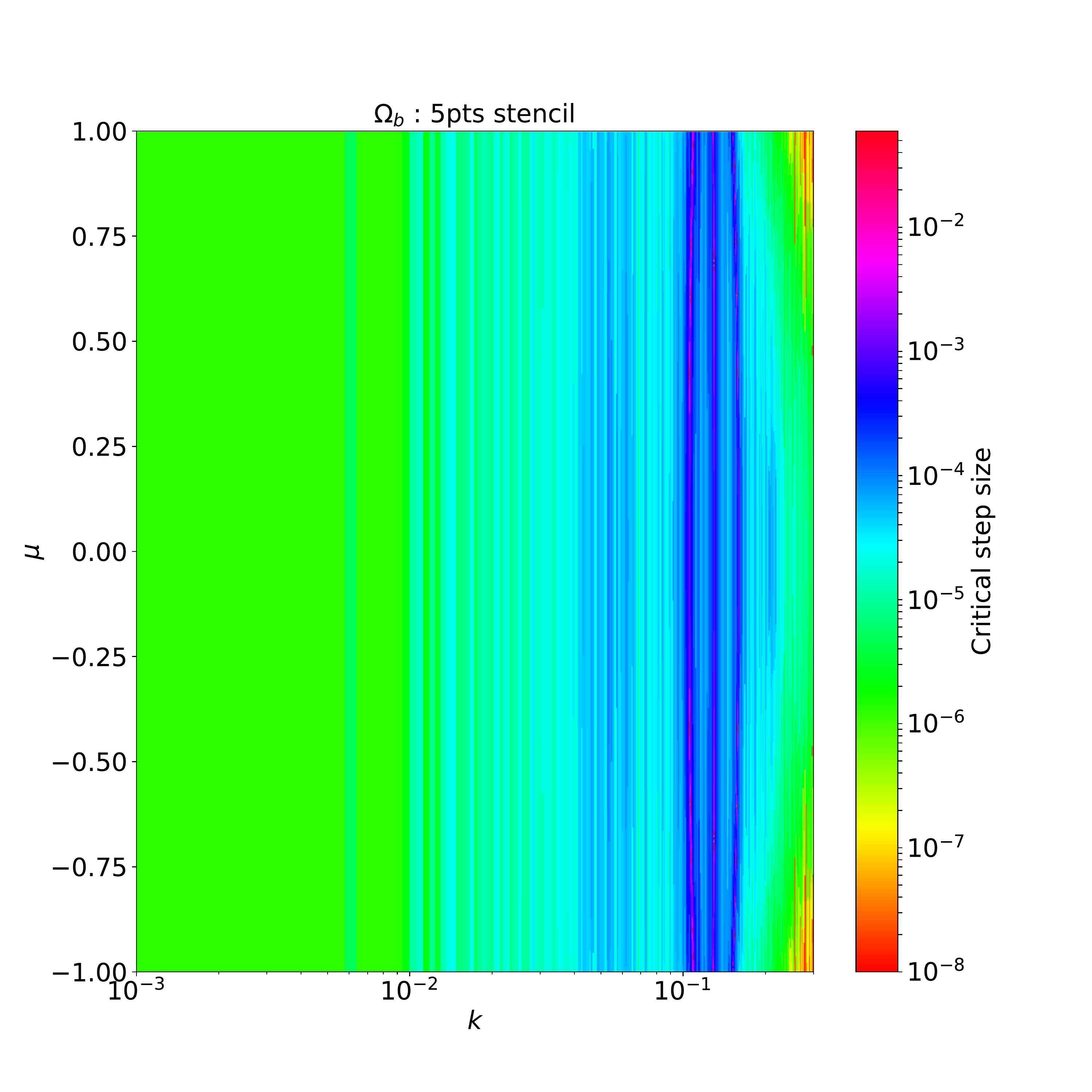} 
\end{array}$
\caption[limit_plot1]{\label{LM_P} Optimal/Limit steps histograms for $\Omega_{\rm b}$ (M1). Upper left panel : optimal steps for the 3 points derivatives ; upper right panel : optimal steps for the 5 points derivatives ; bottom left panel : critical steps for the 3 points derivatives ; bottom right panel : critical steps for the 3 points derivatives.}
\end{figure*}

Considering that $(k, \mu)$ is a $(1000, 1000)$ grid in our forecasts (and we have multiple redshift bins), we preferably need to study convergence in the full grid instead of a specific $(k,\mu)$ value. The upper panels of Figure~\ref{LM_P} summarize the optimal steps corresponding to the convergence level of the square of the derivative of $\ln P(k,\mu)$ with respect to $\Omega_{\rm b}$ (3 points stencil: upper left panel, 5 points stencil: upper right panel) at $z=1.35$. Concerning the 3 points stencil, the optimal step size is located around values of the order of $10^{-2}$ at large scales and $(10^{-2}, 10^{-3})$ at small scales. For the 5 points stencil, it is in the range $10^{-1}$ to $10^{-2}$ at large scales and $10^{-1}$ to $10^{-3}$ at small scales. The bottom panels of Figure~\ref{LM_P} demonstrate the $\Omega_{\rm b}$ critical step size giving a relative error comparable to the elements $(\Omega_{\rm b}, \Omega_{\rm b})$ of the vibration matrix. In other words this is a map of the $\Omega_{\rm b}$ step sizes that can potentially lead to an error of $10\%$ on the final constraints. As we have already seen, the truncation errors are small for the cosmological parameters, and here we show steps that are in the round-off errors regime (see Appendix \ref{sec:definitions} for these and other definitions relevant to the work). The 3 points stencil shows some round-off issues for steps going from $10^{-5}$ (large scales) to $10^{-3}$ (small scales). Concerning the 5 points stencil, the round-off issues occurs for steps from $10^{-6}$ (large scales) to $10^{-3}$ (small scales). The 5 points stencil shows more tolerance than the 3 points stencil. For cosmological parameters choosing a 5 points method with steps around $10^{-2}$ is the safest way to compute derivatives. There is no dependence of the step size on $\mu$ in the M2 model because the derivatives only depend on the matter power spectrum. However, this dependence arises in the M1 model through the Alcock-Paczynski effect. 

\begin{figure*}[t!]
$\begin{array}{rcl}
    \includegraphics[width=0.499\textwidth]{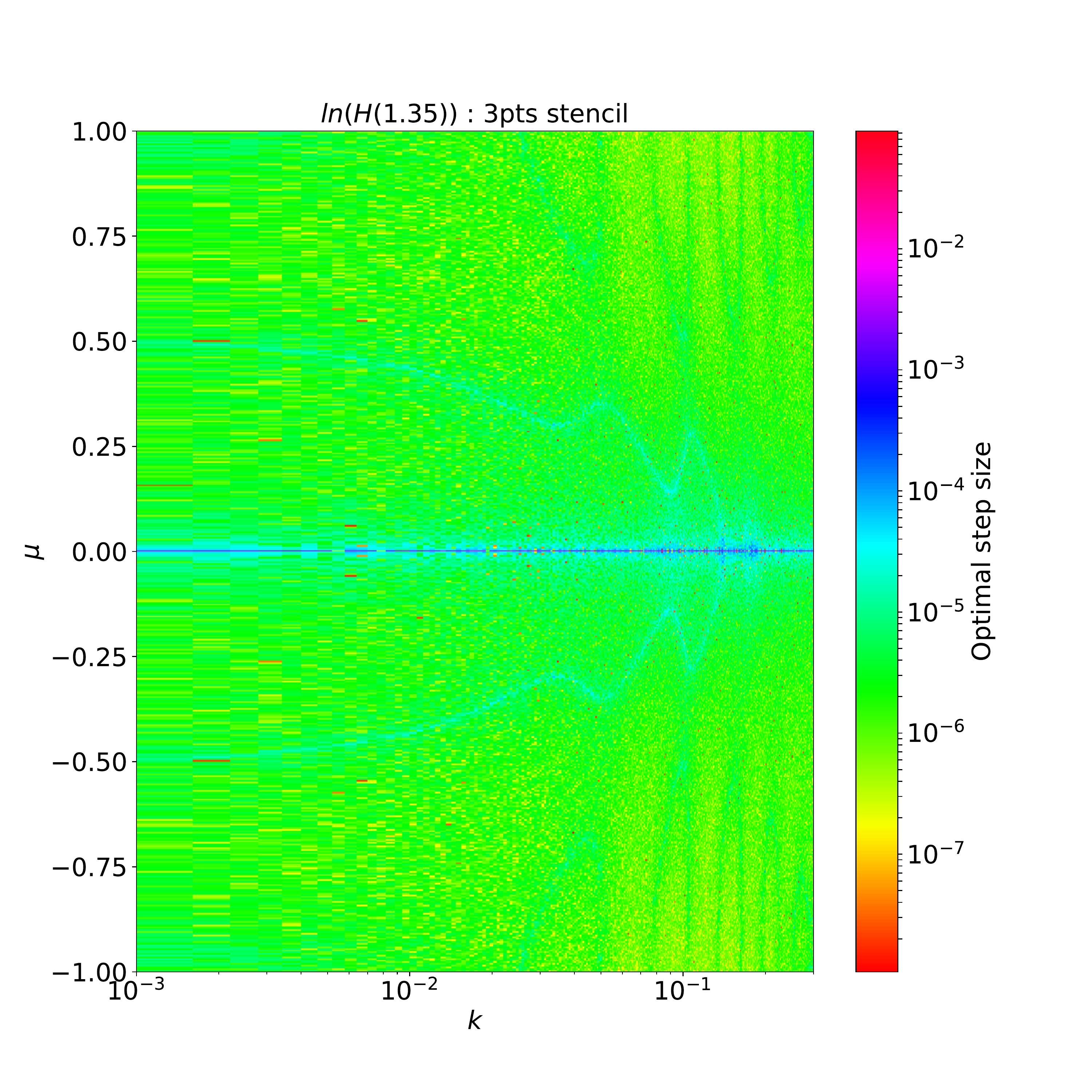} &
    \includegraphics[width=0.499\textwidth]{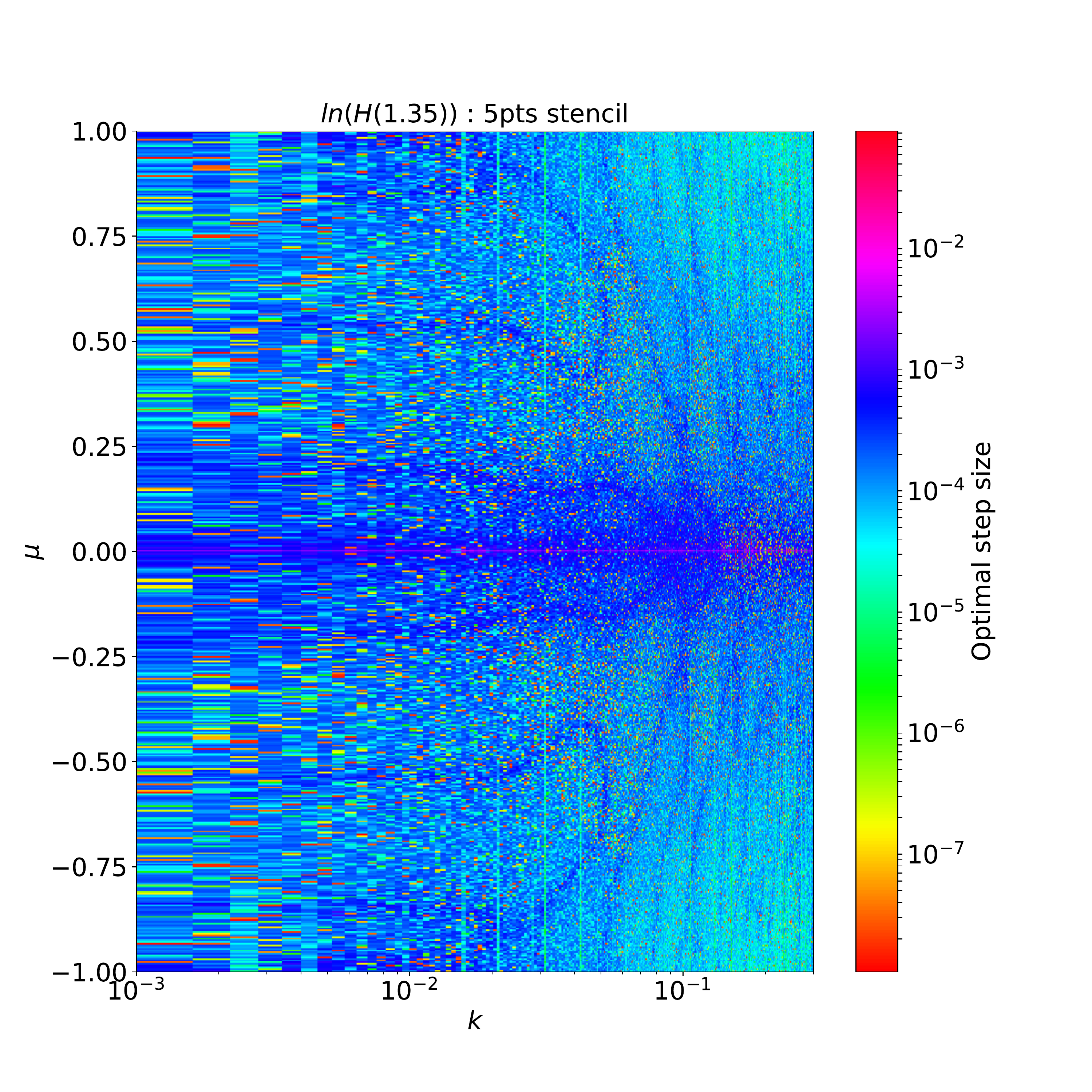} \\
    \includegraphics[width=0.499\textwidth]{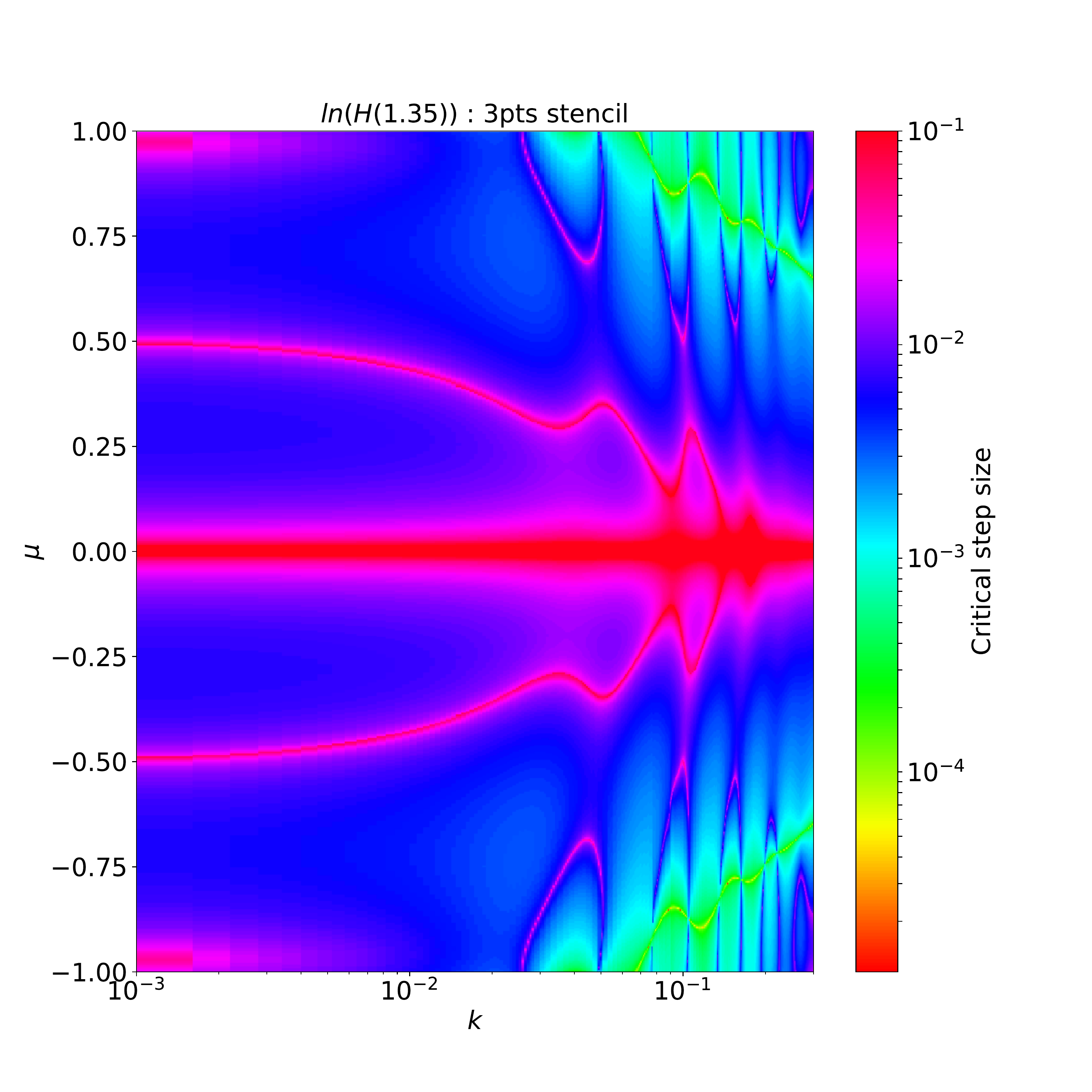} &
    \includegraphics[width=0.499\textwidth]{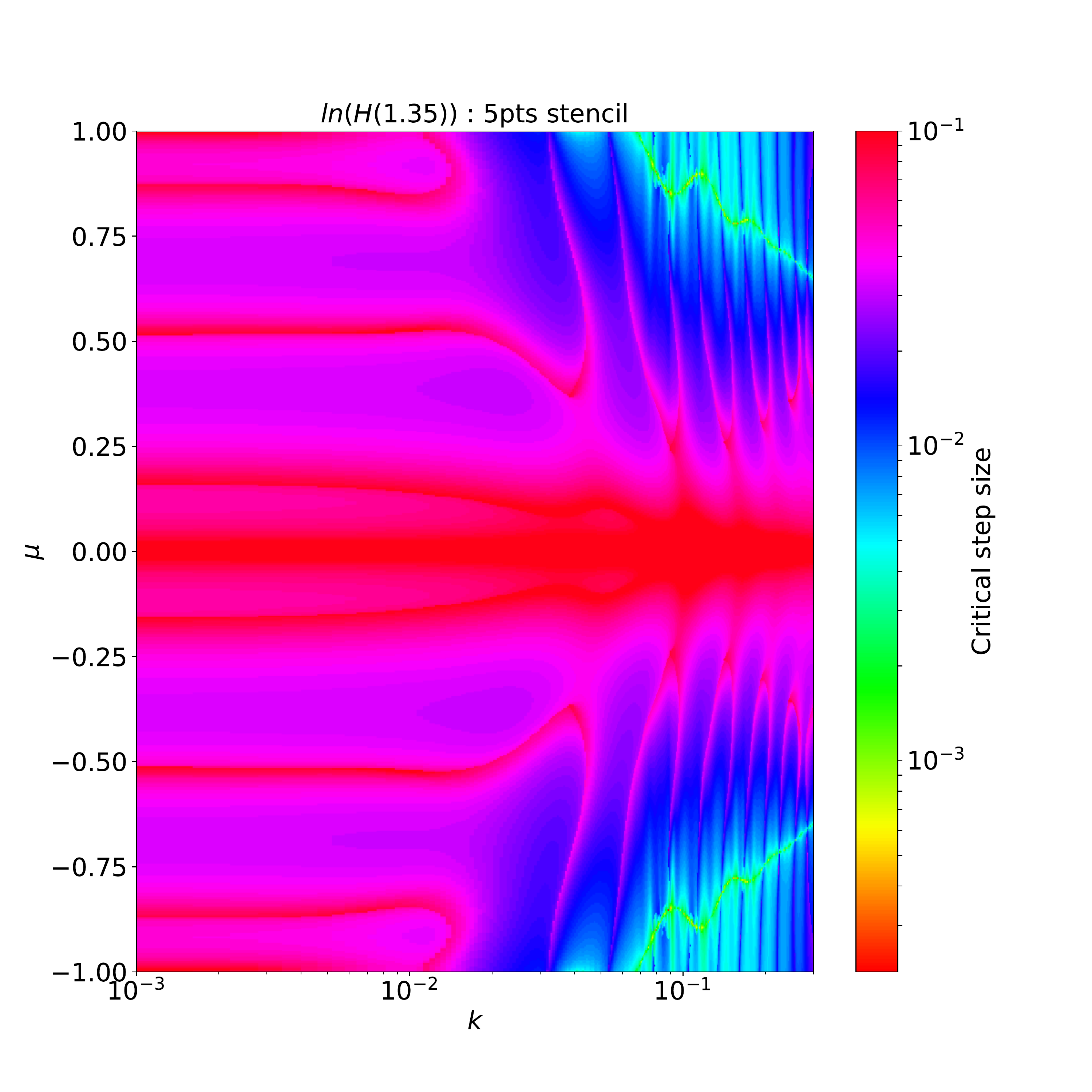} 
\end{array}$
\caption[limit_plot2]{\label{LM_P2} Optimal/Limit steps histograms for ${\rm ln}(H(1.35))$ (M2). Upper left panel: optimal steps for the 3 points derivatives ; upper right panel: optimal steps for the 5 points derivatives ; bottom left panel: critical steps for the 3 points derivatives ; bottom right panel: critical steps for the 3 points derivatives.}
\end{figure*}

Figure \ref{LM_P2} shows the optimal (upper panel) and the critical (bottom panel) step size for ${\rm ln}(H(1.35))$ for the 3 points (left) and 5 points (right) stencil derivative. The optimal step size mainly lies around $10^{-6}$ for the 3 points stencil (even $1.5 \times 10^{-7}$ in certain cells). The 5 points stencil mainly shows convergence from $10^{-5}$ to a few $10^{-4}$. Regarding the critical step size, here we took the truncation error as a limit because of its dominance when we compute derivatives over background quantities. The critical range for the 3 points derivatives extends from $10^{-1}$ to $10^{-4}$, while for the 5 points stencil the range is $10^{-1}$ to a few $10^{-3}$. The lower bound of the critical range is one order of magnitude lower for the 5 points stencil. For the background quantities taking low step sizes is a good option to compute accurate derivatives. The round-off errors remain much less dominant than the truncation errors.   

Ideally one would like to use an optimal  step size for each value of the $k$ and $\mu$. However, from figures 4 and 5 one should choose -- when possible -- a single step size  which will work efficiently over the whole parameter space. This single optimal step size should be below the critical value all over the parameter space. Tables \ref{Steps_T1} and \ref{Steps_T2} summarize the steps chosen for the M1 and M2 approaches using the 5 points method. For cosmological parameters a step between $10^{-1}$ and $10^{-2}$ is enough to compute accurate derivatives. For background quantities we use $10^{-4}$. We added the critical steps and hypercritical steps. The former corresponds to a choice of step size ensuring a precision on the derivatives better than the chosen level while the latter is not expected to lead to the requested precision. The hypercritical steps are the same as the critical steps except for $w_a$ whose step size is twice smaller (M1 approach) in order to increase the numerical noise; we also multiply by a factor $2$ the steps of the background quantities (M2 approach) to increase the truncation errors. The constraints obtained by these $3$ step cases are compared with the MCMC sampling.  

\begin{table}
\centering
\small
\setlength\tabcolsep{1pt}
\begin{tabular}{ccccccccc}
  \hline
  parameter & $\Omega_{\rm b}$ & $h$ & $\Omega_{\rm m}$ & $n_{\rm s}$ & $\Omega_{\rm DE}$ & $w_0$ & $w_a$ & $\sigma_8$ \\
  \hline \hline
  optimal step & 3e-2 & 1e-2 & 2e-2 & 1e-1 & 4e-2 & 2e-2 & 8e-2 & 1e-2 \\ 
  \hline \hline
  critical step & 1e-5 & 1e-1 & 1e-1 & 1e-1 & 1e-1 & 1e-5 & 2e-5 & 1e-12 \\ 
  \hline 
  hypercritical step & 1e-5 & 1e-1 & 1e-1 & 1e-1 & 1e-1 & 1e-5 & 1e-5 & 1e-12 \\ 
  \hline \hline
\end{tabular}
\caption{Step values (approach M1): 5 points derivatives.}
\label{Steps_T1}
\end{table}

\begin{table}
\centering
\small
\setlength\tabcolsep{1pt}
\begin{tabular}{ccccccccc}
  \hline
  parameter & $\omega_{\rm b}$ & $h$ & $\omega_{\rm m}$ & $n_{\rm s}$ & $\ln(D_a(z))$ & $\ln(H(z))$ \\
  \hline \hline
  optimal step & 2e-2 & 1e-2 & 2e-2 & 1e-1 & 1e-4 & 1e-4 \\ 
  \hline 
  critical step & 2e-2 & 1e-2 & 2e-2 & 1e-1 & 1e-2 & 1e-2 \\ 
  \hline 
  hypercritical step & 2e-2 & 1e-2 & 2e-2 & 1e-1 & 2e-2 & 2e-2 \\  
  \hline \hline
\end{tabular}
\caption{Step values (approach M2): 5 points derivatives.}
\label{Steps_T2}
\end{table}

\subsection{Consistency: approach M1 vs approach M2}

\begin{figure*}[t!]
$\begin{array}{rcl}
    \includegraphics[width=0.999\textwidth]{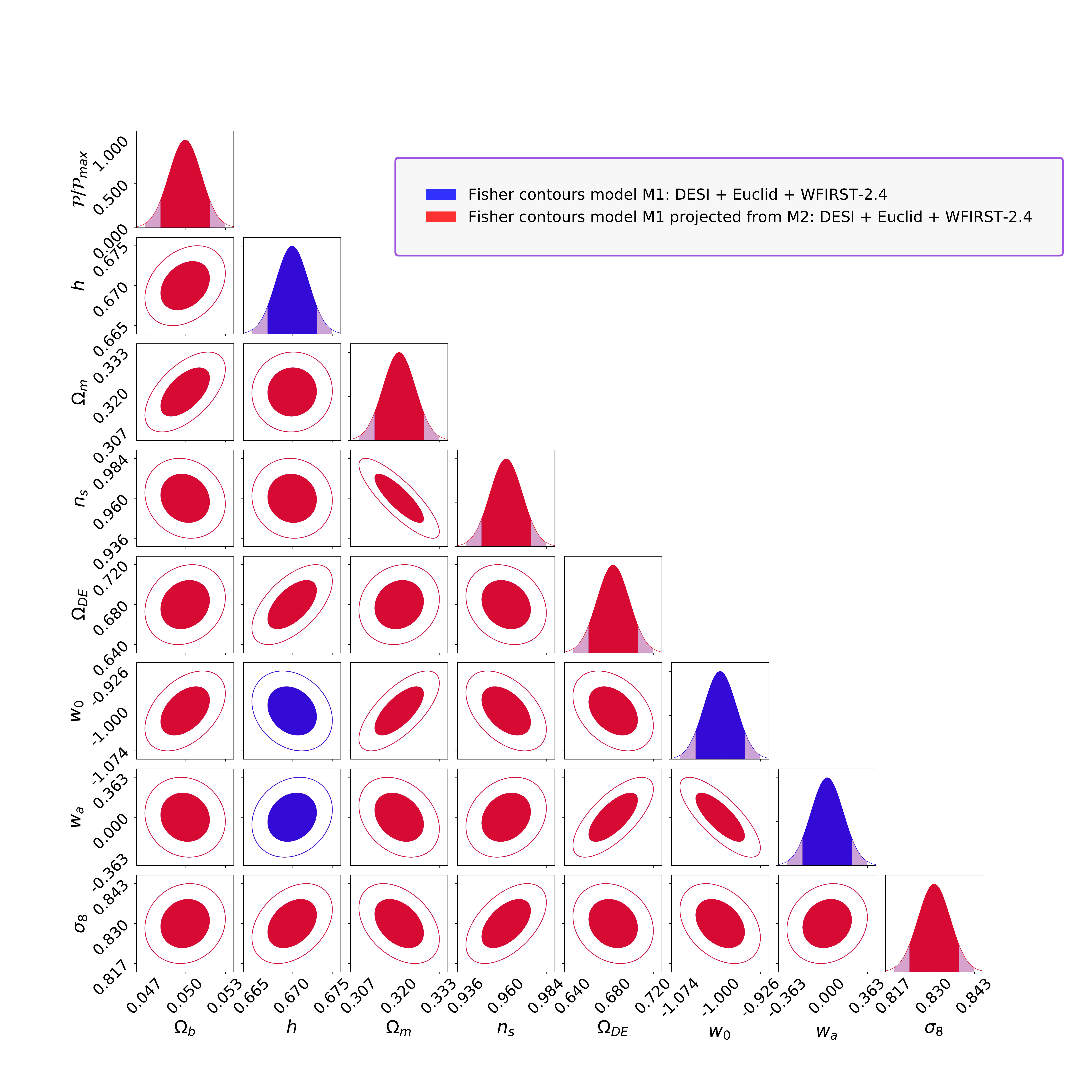} &
\end{array}$
\caption[CST_plot]{\label{CST_M1M2}Fisher approach M1 (blue) contours with optimal step sizes vs Fisher approach M1 (red) with optimal step sizes contours projected from the model M2. Smaller contours are set in front. Here contours are nearly identical, minor numerical differences put one contour in front of the other nearly randomly.}
\end{figure*}

In this section we test the consistency of our results by comparing the two approaches, M1 and M2. This can be achieved by projecting the Fisher matrix from the \itt{M2} parameter space to the \itt{M1} parameter space. Let us consider the initial set of parameters given by the model M2 (these are $P_{\rm initial}$[$\omega_{\rm b}$, $h$, $\omega_{\rm m}$, $n_{\rm s}$, $\ln(D_a(0.125))$, $\ln(H(0.125))$, $\ln(f\sigma_8(0.125))$,  $\ln(b\sigma_8(0.125))$, $P_{\rm shot}(0.125)$, ..., $\ln(D_a(2.55))$, $\ln(H(2.55))$, $\ln(f\sigma_8(2.55))$, $\ln(b\sigma_8(2.55))$, $P_{\rm shot}(2.55)$]) and the final set of parameters (model M1), that is  $P_{\rm final}$[$\Omega_{\rm b}$, $h$, $\Omega_{\rm m}$, $n_{\rm s}$, $\Omega_{\rm DE}$, $w_0$, $w_a$, $\sigma_8$, $\ln(b\sigma_8(0.125))$, $P_{\rm shot}(0.125)$, ..., $\ln(b\sigma_8(2.55))$, $P_{\rm shot}(2.55)$]. We can project the initial Fisher matrix to the final one by simply using the chain rule \citep{wang_dark_2006,wang_designing_2010}. The marginalization over the nuisance parameters can be done before or after the projection. In the latter case they have to be taken into account during the projection:

\begin{equation}
F_{\rm final_{ij}} = \sum_{\alpha \beta} \frac{\partial{P_{\rm initial_{\alpha}}}}{\partial{P_{\rm final_i}}} \left(F_{\rm initial_{\alpha \beta}} \right) \frac{\partial{P_{\rm initial_\beta}}}{\partial{P_{\rm final_j}}} \, ,
\label{Eq9}
\end{equation}

\noindent where indices $i$ and $j$ run over all unique pair of parameters corresponding to the final model (we build the other half of the final Fisher matrix by computing the symmetric (by mirroring over the diagonal). Indices $\alpha$ and $\beta$ run over all parameter pairs corresponding to the M1 approach. We summarize the constraints relative errors obtained from the comparison between the standard M1 approach computation against the constraints after projection from the parameters introduced in the M2 approach in Table \ref{M1vsM2T}. Both covariances are marginalized over the nuisance parameters. We find that the two methods are in very good  agreement, with the relative errors not exceeding $\sim 0.02\%$. The corresponding contours are shown in Figure~\ref{CST_M1M2}. As we can see, they are nearly perfectly superimposed. Therefore, the final Fisher matrices are highly consistent in the two approaches.   

\begin{table}
\centering
\small
\setlength\tabcolsep{1pt}
\begin{tabular}{ccccccccc}
  \hline \hline
  parameter & $\Omega_{\rm b}$ & $h$ & $\Omega_{\rm m}$ & $n_{\rm s}$ \\
  \hline
  relative error (\%) & 1.857e-3 & 2.332e-3 & -2.176e-2 & 2.155e-2\\ 
  \hline 
  parameter & $\Omega_{\rm DE}$ & $w_0$ & $w_a$ & $\sigma_8$ \\
  \hline 
  \textbf{relative error (\%)} & -1.137e-3 & -7.339e-3 & -1.573e-3 & -5.706e-3 \\  
  \hline \hline
\end{tabular}
\caption{Relative errors on the cosmological parameter constraints between the M1 and M2 approaches (both with optimal step sizes) after projection to the $\theta_{\rm cosmo}$ parameter set. In both cases, the covariance matrices are marginalized over the nuisance parameters.}
\label{M1vsM2T}
\end{table}

\subsection{Fisher matrix and MCMC comparison}

\begin{figure*}[t!]
$\begin{array}{rcl}
    \includegraphics[width=0.999\textwidth]{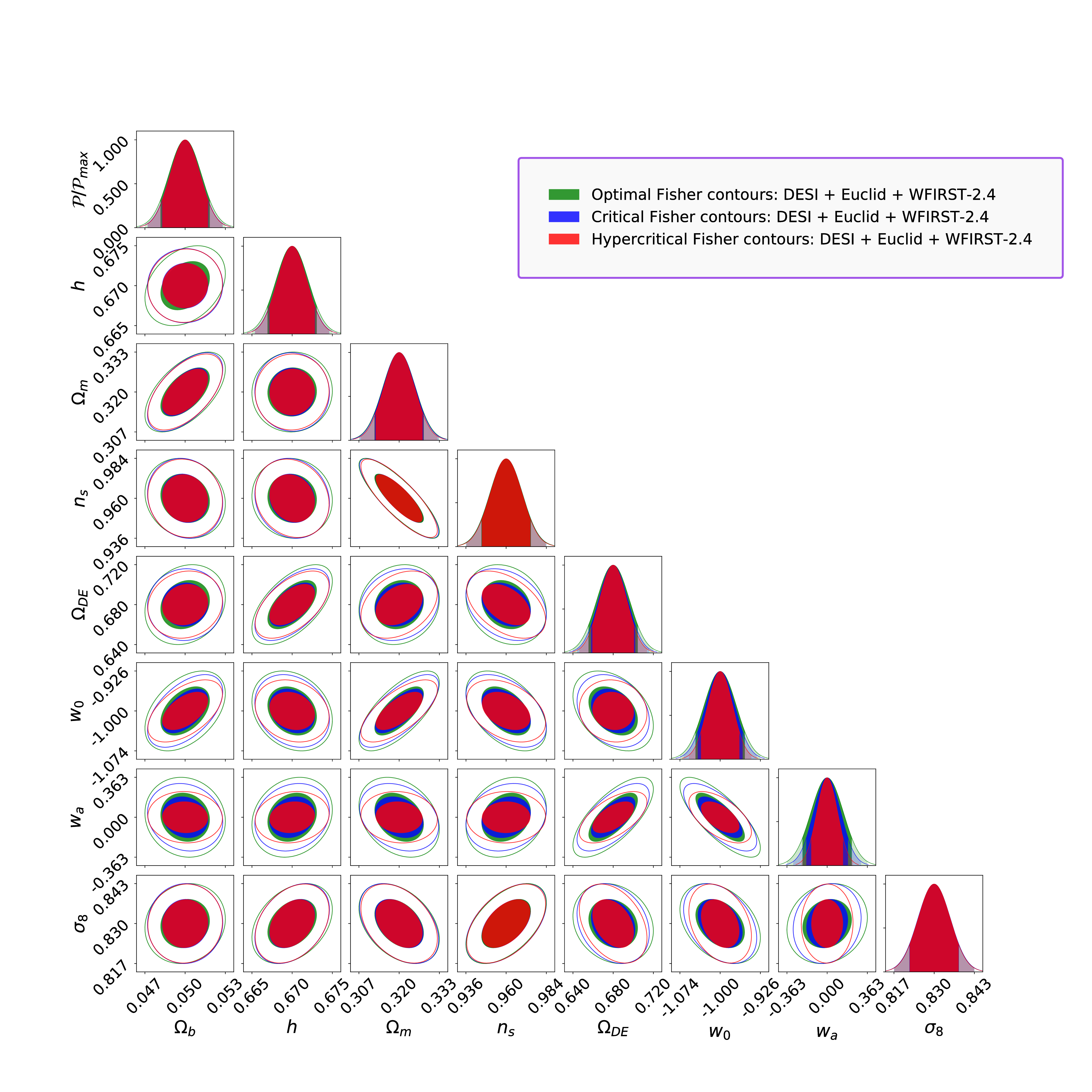} &
\end{array}$
\caption[Ctr_plot1]{\label{CT_P1} Fisher contours:  M1 approach, for the optimal (green), critical (blue), and hypercritical (red) steps. The contours are significantly affected when using inappropriate step size, although the 1D likelihoods remains pretty stable.} 
\end{figure*}

\begin{figure*}[t!]
$\begin{array}{rcl}
    \includegraphics[width=0.999\textwidth]{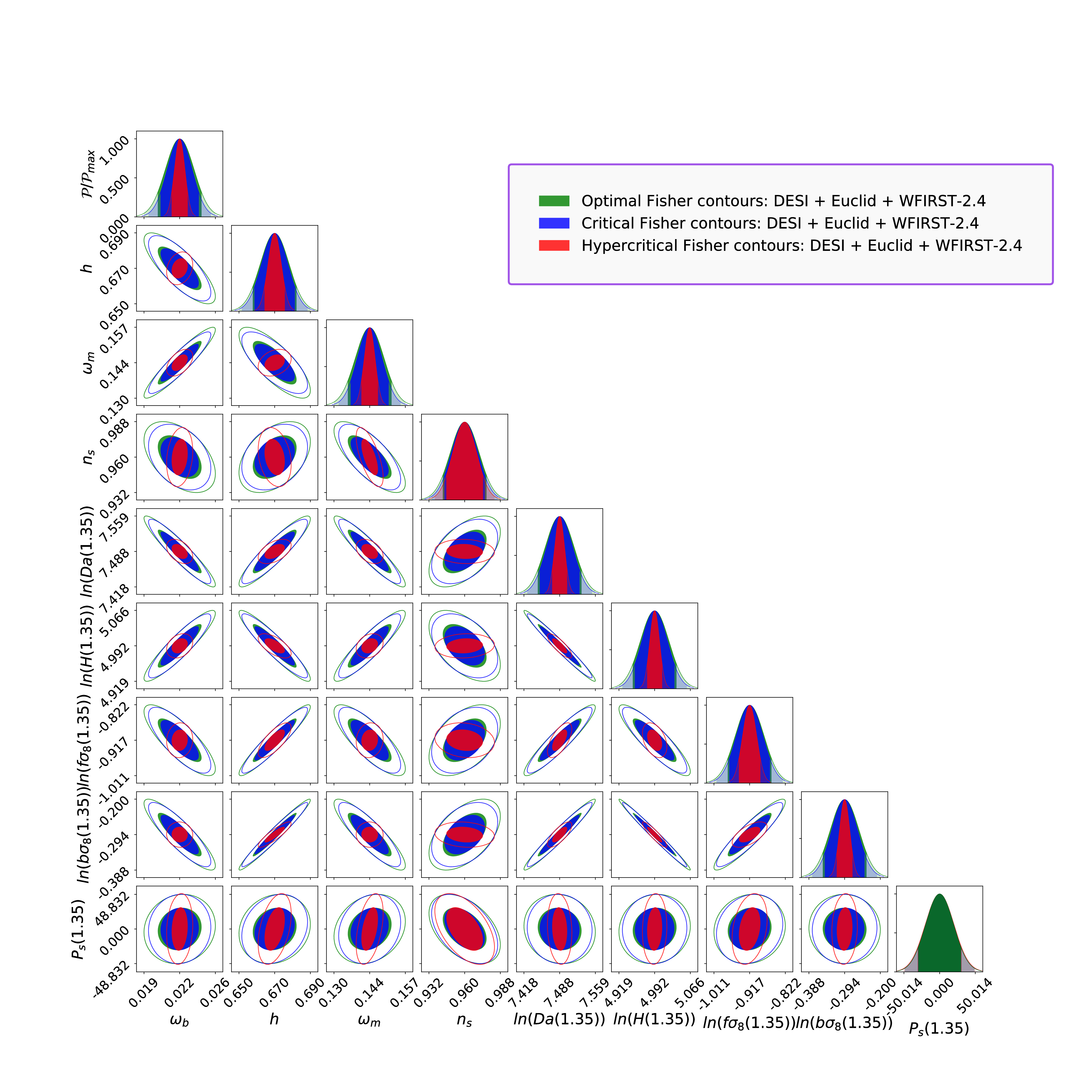} &
\end{array}$
\caption[Ctr_plot2]{\label{CT_P2} Fisher contours: M2 approach, for the optimal (green), critical (blue), and hypercritical (red) steps. In this case, the use of critical step size leads to contours close the ones obtained with the optimal step size, while the hypercritical (twice the critical one) leads most of the contours to be incorrect, as well as most of the 1D likelihood.}
\end{figure*}

The relative errors between the MCMC and the Fisher constraints are presented in Table~\ref{MCMC_C1}. With the optimal steps the relative error is lower than $5.5\%$ for both models. With the critical steps, the $w_a$ relative error exceeds $10\%$ (model M1). The other parameters have errors ranging from $2 \times 10^{-3}$ to $\sim 6\%$. In the second approach every step agrees well with the MCMC with a highest relative error of $9.118\%$. Choosing hypercritical steps leads to significant disagreement with the MCMC constraints. In the M1 approach, three cosmological parameters ($\Omega_{\rm DE}$, $w_0$ and $w_a$) and one nuisance parameter ($\ln(b\sigma_8(0.375))$) do not fulfill the $10\%$ agreement requirement level. The relative error on $w_a$ reaches $34\%$. In the M2 approach, the results are even worse, with most of the parameters having errors above $50\%$. Figures~\ref{CT_P1} and \ref{CT_P2} show the Fisher contours using the optimal, critical, and hypercritical steps set. The covariances are marginalized over the nuisance parameters. For convenience, we only show the cosmological parameters for the M1 approach, and the cosmological as well as one redshift bin ($z=1.35$) for the approach M2. In the M1 approach, Figure~\ref{CT_P1}, the $\Omega_{\rm b}$, $h$, $\Omega_{\rm m}$, $n_{\rm s}$ and $\sigma_8$ constraints are very stable for the 3 steps set. The $\Omega_{\rm DE}$ parameter shows a slight difference between the 3 steps set. The dark energy parameters $w_0$ and $w_a$ show more appreciable differences, especially $w_a$. In the M2 approach (Figure~\ref{CT_P2}) we can only see a few differences between the optimal and the critical steps set. However the hypercritical steps set gives very different results on all parameters except $n_{\rm s}$ and $P_{\rm shot}(z))$. The constraints are heavily overestimated (the uncertainties are too small) and fill around one third of the two other surface contours. We emphasise here that the only difference between the critical steps and the hypercritical steps is a factor of $2$ on the steps for only one parameter in the M1 approach and two parameters on the M2 approach. The transition from the stable derivatives to the unstable derivatives is very sharp.  

\begin{table}
\centering
\small
\setlength\tabcolsep{1pt}
\begin{tabular}{|c|c|c|c|}
  \hline
  \multicolumn{4}{|c|}{Relative error MCMC vs Fisher matrix: model M1 ($\%$)} \\
  \hline
  $parameters$ & $optimal$ & $critical$ & $h critical$ \\
  \hline
  $\Omega_{\rm b}$ & $-1.201$ & $5.302$ & $6.143$ \\
  \hline
  $h$ & $-4.059$ & $3.467$ & $4.351$ \\
  \hline
  $\Omega_{\rm m}$ & $-1.999$ & $-0.294$ & $3.013$ \\
  \hline
  $n_{\rm s}$ & $-1.759$ & $-1.790$ & $0.146$ \\
  \hline
  $\Omega_{\rm DE}$ & $-4.723$ & $5.899$ & $11.916$ \\
  \hline
  $w_0$ & $-3.214$ & $6.133$ & $19.704$ \\
  \hline
  $w_a$ & $-5.183$ & $13.365$ & $34.022$ \\ 
  \hline
  $\sigma_8$ & $-3.341$ & $-3.305$ & $-2.256$ \\
  \hline
  $\ln(b\sigma_8(0.125))$ & $-3.418$ & $1.033$ & $5.306$ \\
  \hline
  $P_{\rm shot}(0.125))$ & $-3.523$ & $-3.223$ & $-2.736$ \\
  \hline
  $\ln(b\sigma_8(0.375))$ & $-3.067$ & $4.248$ & $11.174$ \\
  \hline
  $P_{\rm shot}(0.375))$ & $-2.804$ & $-1.042$ & $1.760$ \\
  \hline
  $\ln(b\sigma_8(0.6))$ & $-2.404$ & $0.551$ & $3.896$ \\
  \hline
  $P_{\rm shot}(0.6))$ & $-3.692$ & $-3.563$ & $-3.249$ \\
  \hline
  $\ln(b\sigma_8(0.8))$ & $-3.492$ & $1.258$ & $5.734$ \\
  \hline
  $P_{\rm shot}(0.8))$ & $-2.419$ & $-2.177$ & $-1.063$ \\
  \hline
  $\ln(b\sigma_8(1.05))$ & $-4.591$ & $0.531$ & $4.042$ \\
  \hline
  $P_{\rm shot}(1.05))$ & $-3.411$ & $-2.883$ & $-2.725$ \\
  \hline
  $\ln(b\sigma_8(1.35))$ & $-3.732$ & $0.679$ & $3.312$ \\
  \hline
  $P_{\rm shot}(1.35))$ & $-3.149$ & $-2.626$ & $-2.643$ \\
  \hline
  $\ln(b\sigma_8(1.65))$ & $-3.487$ & $0.002$ & $2.108$ \\
  \hline
  $P_{\rm shot}(1.55))$ & $-3.866$ & $-3.480$ & $-3.592$ \\
  \hline
  $\ln(b\sigma_8(1.95))$ & $-2.445$ & $0.388$ & $2.568$ \\
  \hline
  $P_{\rm shot}(1.95))$ & $-4.662$ & $-3.730$ & $-3.730$ \\
  \hline
  $\ln(b\sigma_8(2.25))$ & $-1.965$ & $-0.012$ & $1.236$ \\
  \hline
  $P_{\rm shot}(2.25))$ & $-0.671$ & $-0.378$ & $-0.427$ \\
  \hline
  $\ln(b\sigma_8(2.55))$ & $-4.628$ & $-3.197$ & $-2.449$ \\ 
  \hline
  $P_{\rm shot}(2.55))$ & $-2.801$ & $-2.566$ & $-2.605$ \\
  \hline
\end{tabular}
\caption{MCMC vs Fisher relative errors for the model M1. hcritical stands for the hypercritical steps choice described in the text.}
\label{MCMC_C1}
\end{table} 

\begin{table}
\centering
\small
\setlength\tabcolsep{1pt}
\begin{tabular}{|c|c|c|c|}
  \hline 
  \multicolumn{4}{|c|}{Relative error MCMC vs Fisher: model M2 ($\%$)} \\
  \hline
  $parameters$ & $optimal$ & $critical$ & $h critical$ \\
  \hline
  $\omega_{\rm b}$ & $-4.044$ & $9.118$ & $61.864$ \\
  \hline
  $h$ & $-5.467$ & $2.818$ & $51.171$ \\
  \hline
  $\omega_{\rm m}$ & $-5.183$ & $8.614$ & $60.223$ \\
  \hline
  $n_{\rm s}$ & $-2.302$ & $5.236$ & $14.492$ \\
  \hline
  $\ln(D_a(0.125))$ & $-4.134$ & $2.398$ & $30.746$ \\
  \hline
  $\ln(H(0.125))$ & $-4.108$ & $2.378$ & $31.556$ \\
  \hline
  $\ln(f\sigma_8(0.125))$ & $-2.270$ & $1.044$ & $15.818$ \\
  \hline
  $\ln(b\sigma_8(0.125))$ & $-4.901$ & $3.204$ & $44.078$ \\
  \hline
  $P_{\rm shot}(0.125))$ & $-3.641$ & $-3.351$ & $-4.703$ \\
  \hline
  $\ln(D_a(0.375))$ & $-5.320$ & $3.814$ & $51.556$ \\
  \hline
  $\ln(H(0.375))$ & $-5.070$ & $3.932$ & $50.975$ \\
  \hline
  $\ln(f\sigma_8(0.375))$ & $-3.949$ & $0.883$ & $23.873$ \\
  \hline
  $\ln(b\sigma_8(0.375))$ & $-5.460$ & $3.955$ & $56.284$ \\
  \hline
  $P_{\rm shot}(0.375))$ & $-3.494$ & $-1.875$ & $-1.224$ \\
  \hline
  $\ln(D_a(0.6))$ & $-3.538$ & $3.266$ & $33.064$ \\
  \hline
  $\ln(H(0.6))$ & $-3.888$ & $3.262$ & $36.254$ \\
  \hline
  $\ln(f\sigma_8(0.6))$ & $-3.062$ & $3.093$ & $32.527$ \\
  \hline
  $\ln(b\sigma_8(0.6))$ & $-4.265$ & $3.307$ & $39.351$ \\
  \hline
  $P_{\rm shot}(0.6))$ & $-1.530$ & $-1.374$ & $-2.117$ \\
  \hline
  $\ln(D_a(0.8))$ & $-4.993$ & $4.691$ & $56.769$ \\
  \hline
  $\ln(H(0.8))$ & $-4.845$ & $4.799$ & $57.621$ \\
  \hline
  $\ln(f\sigma_8(0.8))$ & $-5.187$ & $3.423$ & $52.354$ \\
  \hline
  $\ln(b\sigma_8(0.8))$ & $-5.230$ & $4.365$ & $57.019$ \\
  \hline
  $P_{\rm shot}(0.8))$ & $-1.685$ & $-0.413$ & $-39.840$ \\
  \hline
  $\ln(D_a(1.05))$ & $-5.359$ & $4.207$ & $64.246$ \\
  \hline
  $\ln(H(1.05))$ & $-5.483$ & $4.532$ & $64.077$ \\
  \hline
  $\ln(f\sigma_8(1.05))$ & $-5.123$ & $2.736$ & $50.217$ \\
  \hline
  $\ln(b\sigma_8(1.05))$ & $-5.658$ & $4.088$ & $62.095$ \\
  \hline
  $P_{\rm shot}(1.05))$ & $-2.900$ & $-2.942$ & $-4.709$ \\
  \hline
  $\ln(D_a(1.35))$ & $-5.119$ & $4.441$ & $64.000$ \\
  \hline
  $\ln(H(1.35))$ & $-4.976$ & $5.192$ & $64.010$ \\
  \hline
  $\ln(f\sigma_8(1.35))$ & $-5.474$ & $1.986$ & $48.206$ \\
  \hline
  $\ln(b\sigma_8(1.35))$ & $-5.302$ & $4.551$ & $62.316$ \\
  \hline
  $P_{\rm shot}(1.35))$ & $-3.530$ & $-3.712$ & $-60.378$ \\
  \hline
  $\ln(D_a(1.65))$ & $-4.991$ & $4.625$ & $60.693$ \\
  \hline
  $\ln(H(1.65))$ & $-5.091$ & $5.129$ & $60.626$ \\
  \hline
  $\ln(f\sigma_8(1.65))$ & $-4.595$ & $2.038$ & $40.691$ \\
  \hline
  $\ln(b\sigma_8(1.65))$ & $-5.400$ & $4.585$ & $60.985$ \\
  \hline
  $P_{\rm shot}(1.65))$ & $-3.073$ & $-3.430$ & $-6.134$ \\
  \hline
  $\ln(D_a(1.95))$ & $-4.837$ & $2.849$ & $54.337$ \\
  \hline
  $\ln(H(1.95))$ & $-4.953$ & $4.029$ & $54.718$ \\
  \hline
  $\ln(f\sigma_8(1.95))$ & $-4.187$ & $0.337$ & $30.715$ \\
  \hline
  $\ln(b\sigma_8(1.95))$ & $-5.276$ & $3.921$ & $58.407$ \\
  \hline
  $P_{\rm shot}(1.95))$ & $-2.426$ & $-3.054$ & $-7.641$ \\
  \hline
  $\ln(D_a(2.25))$ & $-3.534$ & $3.803$ & $41.734$ \\
  \hline
  $\ln(H(2.25))$ & $-3.913$ & $4.231$ & $42.031$ \\
  \hline
  $\ln(f\sigma_8(2.25))$ & $-1.690$ & $0.720$ & $14.056$ \\
  \hline
  $\ln(b\sigma_8(2.25))$ & $-4.566$ & $4.405$ & $51.277$ \\
  \hline
  $P_{\rm shot}(2.25))$ & $-1.741$ & $-2.232$ & $-6.015$ \\
  \hline
  $\ln(D_a(2.55))$ & $-2.590$ & $3.662$ & $31.678$ \\
  \hline
  $\ln(H(2.55))$ & $-3.687$ & $3.179$ & $32.003$ \\
  \hline
  $\ln(f\sigma_8(2.55))$ & $0.659$ & $1.863$ & $7.551$ \\
  \hline
  $\ln(b\sigma_8(2.55))$ & $-4.272$ & $4.030$ & $44.289$ \\
  \hline
  $P_{\rm shot}(2.55))$ & $-1.889$ & $-2.294$ & $-5.557$ \\
  \hline
\end{tabular}
\caption{MCMC vs Fisher relative errors (in percentage) for the M2 approach; \emph{hcritical} refers to the hypercritical steps choice described in the main text.}
\label{MCMC_C2}
\end{table} 

\begin{figure*}[t!]
$\begin{array}{rcl}
    \includegraphics[width=0.999\textwidth]{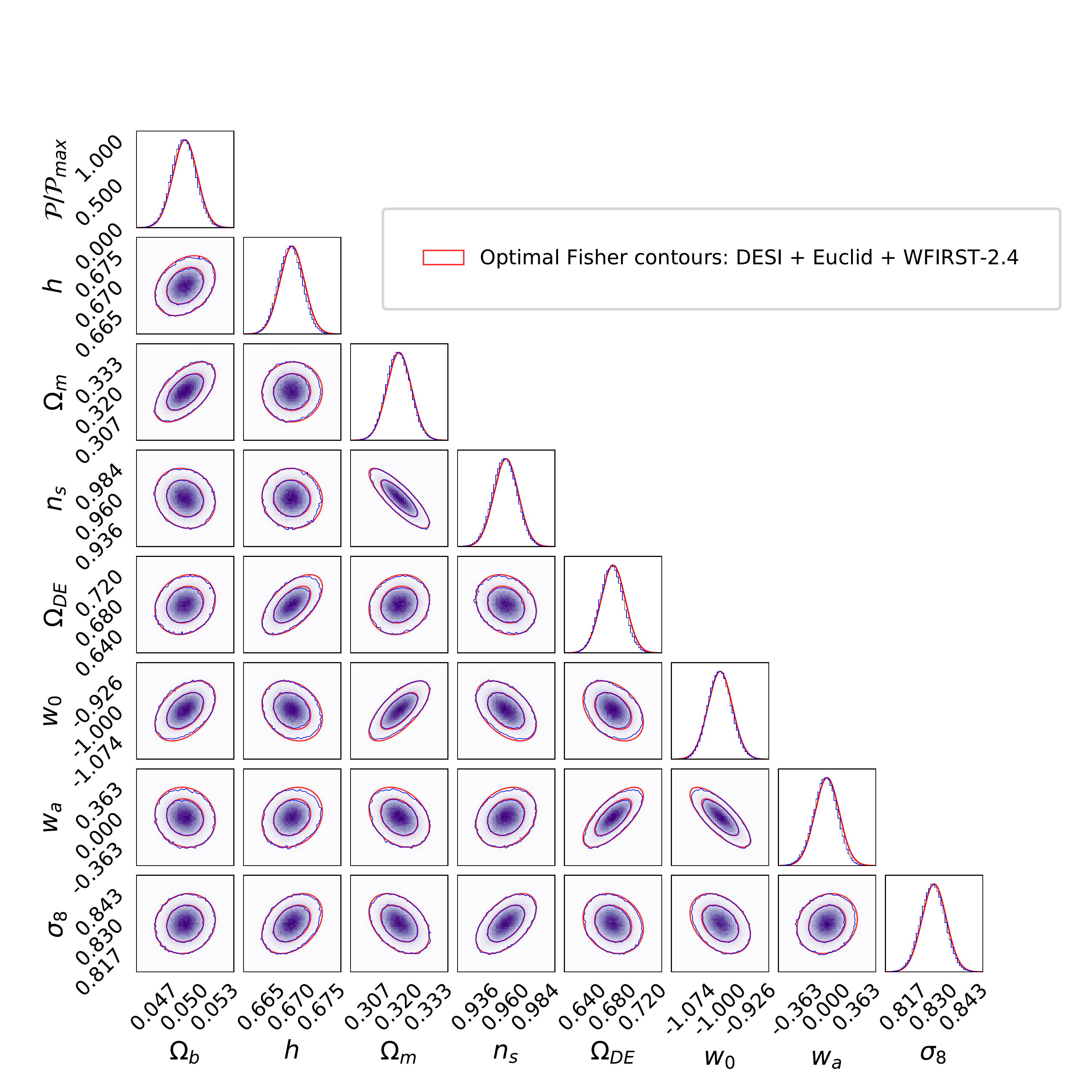} &
\end{array}$
\caption[Ctr_plot3]{\label{CT_P3} MCMC vs Fisher optimal (red) contours : model M1.}
\end{figure*}

\begin{figure*}[t!]
$\begin{array}{rcl}
    \includegraphics[width=0.999\textwidth]{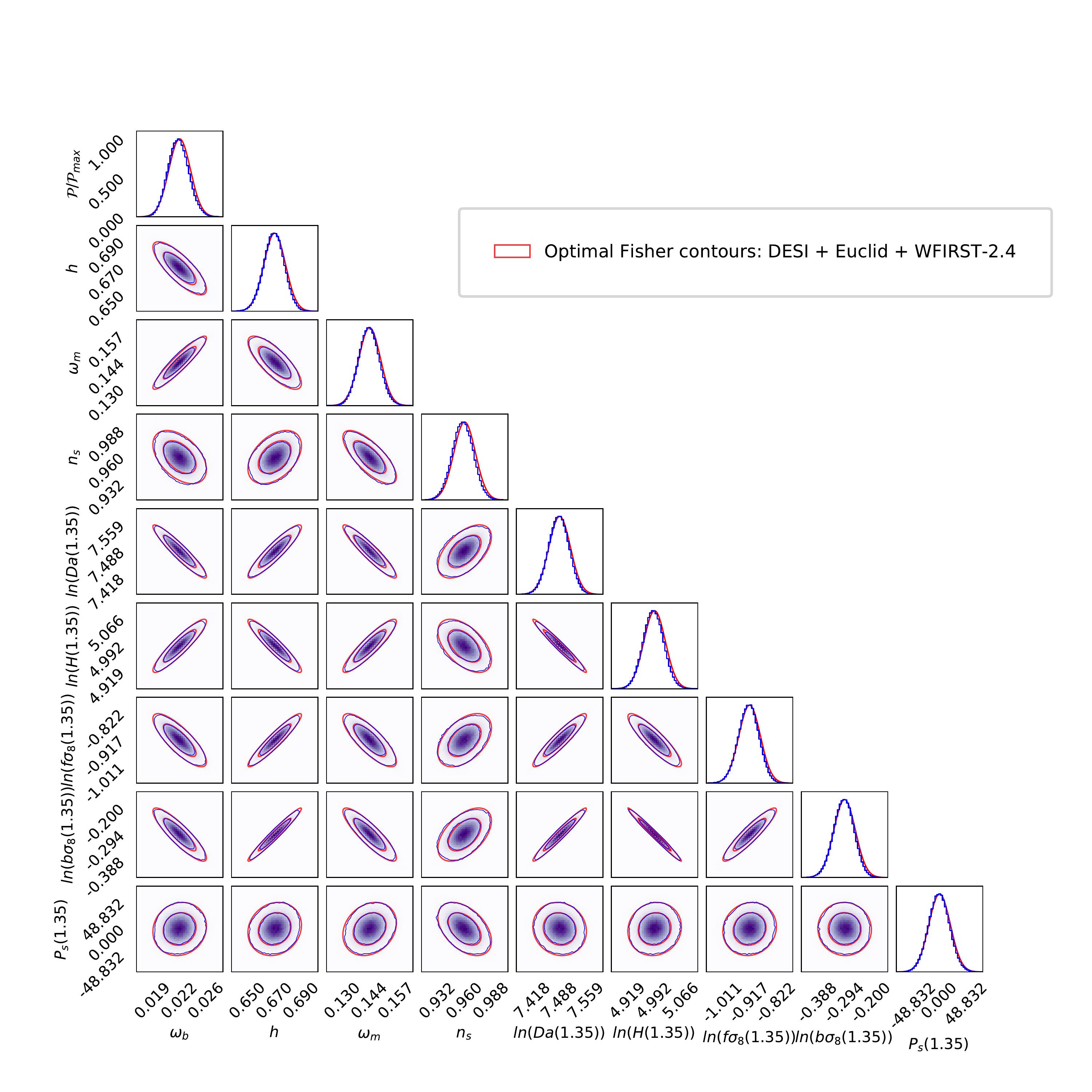} &
\end{array}$
\caption[Ctr_plot4]{\label{CT_P4} MCMC vs Fisher optimal (red) contours : approach M2.}
\end{figure*}

Checking the Fisher matrix derived constraints against the MCMC can give valuable information regarding the precision of the Fisher matrix estimation. However, this is not enough to validate the full procedure. Indeed, two multi dimensional inference methods can result in the same constraints for each individual parameter, but with different contours orientation. In order to ensure that this is not the case, the full posterior distribution has to be compared so one can check that all the contours orientations are similar. Figures~\ref{CT_P3} and \ref{CT_P4} present the MCMC posterior distributions against the Fisher matrix ones (red lines) using the optimal steps settings. All the Fisher contours have the same orientation as the MCMC contours. Thus, the full procedure is validated. It is important to note that, fortunately, using one optimal step for all ($k$, $\mu$, $z$) values is sufficient to have a very good level of agreement with the MCMC. That is to say, no adaptive derivatives approach is required. Moreover, the main differences between the Fisher and the MCMC contours come from slight non-Gaussianities on the posterior distribution. On some likelihood histograms in Figures~\ref{CT_P3} and \ref{CT_P4} we can see that the posterior from the MCMC follows ``perfectly'' one side of the Fisher, but not the other one (for instance $h$ in the M1 approach). This feature is less visible in the M2 approach likely due to the high number of accepted points in the MCMC chain, but it is still there, for example in the case of $n_{\rm s}$. 

\section{Conclusions} 
\label{SEC5}

We have investigated how to perform a comprehensive and robust validation of Fisher matrix forecasts, using the combination of three Stage IV spectroscopic surveys. We built vibration matrices for two different approaches for a fiducial cosmological model: a non flat $\Lambda$CDM extended model where the dark energy equation of state follows the CPL parameterization which has two free variables: $w_0$ and $w_a$. In the first approach (M1) the Fisher matrix is obtained directly for the cosmological parameters while in the second  approach (M2) the background quantities are varied independently from the shape parameters, resulting in a much larger matrix; the final constraints are projected on the same parameter space for both approaches. We also found that the larger Fisher matrix is globally more sensitive to perturbations (as reflected by its high condition number). For instance the Hubble parameter $h$ is two orders of magnitude more sensitive in the second approach. 

We studied the stability of the numerical derivatives over the cosmological parameters and found that the truncation error is particularly small within the step ranges considered. Cosmological parameters remain stable in the step range from $1 \times 10^{-1}$ to $3 \times 10^{-4}$. Lower step values increase the round-off errors and result in higher total errors. Taking ``large'' steps is the safest way to perform the derivatives with respect to the cosmological parameters. The background quantities behave differently. The round-off errors remain small in all the step ranges considered. However steps larger than $1 \times 10^{-2}$ for a five points stencil derivative give unstable results. Taking ``large'' steps for the background quantities is not safe because the differentiated function can oscillate on a scale smaller than the step size; this is exactly what one must avoid when computing the numerical derivatives. The convergence tests with respect to the 7 points stencil derivatives are performing better for the 5 points stencil (the convergence is reached between one and two orders of magnitude before the 3 points stencil case), and similarly for the convergence level. The safe steps window range of the 5 points stencil is then wider. 

Comparison with the MCMC sampling shows the efficiency of the Fisher formalism even with non adaptive derivatives (at least in the case where the posterior distribution is close to Gaussian). Although the optimal steps change with $k$ and $\mu$, choosing a unique step size for all the parameters does not produce appreciably larger errors that would significantly alter the forecasts. Furthermore, we considered a $1000 \times 1000$ ($k$, $\mu$) grid which, multiplied by the number of redshift bins, amounts to computing $10^7$ derivatives per parameter. One of the reasons for the stability comes from the fact that the contribution 
of the large scales on the derivatives is very small compared to the small scales contribution. Thus, taking optimal steps from high $k$ values remains the best option. We have also seen that the transition between stable and unstable contours is quite sharp. Changing a step size by a factor of $2$ towards the wrong direction can give unstable contours. 

Identifying and explaining the derivatives behaviour and the effects on the constraints is difficult without the tests presented. Our strategy allows for a comprehensive and robust validation of the constraints derived from the Fisher matrix formalism through a numerical method.   

\section*{Acknowledgements}
\addcontentsline{toc}{section}{Acknowledgements}
SC is funded by \textsc{miur} through Rita Levi Montalcini project `\textsc{prometheus} -- Probing and Relating Observables with Multi-wavelength Experiments To Help Enlightening the Universe's Structure'. SC acknowledges support from the `Departments of Excellence 2018-2022' Grant (L.\ 232/2016) awarded by the Italian Ministry of Education, University and Research (\textsc{miur}). KM's part of this research was carried out at the Jet Propulsion Laboratory, California Institute of Technology, under a contract with the National Aeronautics and Space Administration (80NM0018D0004). IT acknowledges support from the Spanish Ministry of Science, Innovation and Universities through grant ESP2017-89838-C3-1-R, and the H2020 programme of the European Commission through grant 776247. AP is a UK Research and Innovation Future Leaders Fellow, grant MR/S016066/1.

\nocite{*}
\bibliographystyle{aa}
\bibliography{Safir_Y-C}

\appendix
\numberwithin{equation}{section}
\makeatletter
\newcommand{\section@cntformat}{Appendix \thesection:\ }
\makeatother

\section{The Metropolis-Hastings sampler} \label{sec:Metropolis-Hastings}

Here we describe the steps to sample the parameter space from the \texttt{SpecSAF} MCMC module:

\begin{enumerate}
\item We initialize each parameter at its fiducial value and compute the corresponding observed power spectrum (Equation~\ref{Eq4}).
\item We draw a new parameter set from a multivariate normal distribution (mean = current parameter set) and compute the new observed power spectrum. The multivariate normal distribution is the one given by the inverse of the Fisher matrix for the same specifications (i.e., the covariance matrix).
\item We perform a $\chi^2 $ test in order to get the sum of the square errors :
\begin{equation}
\chi^2_{\rm new} = \sum\limits_{z,k,\mu} \frac{(\ln P_{\rm obs,new}(z,k,\mu) - \ln P_{\rm obs,previous}(z,k,\mu))^2}{\sigma_p(z,k,\mu)^2} \, ,
\label{A1}
\end{equation}

\noindent with $\sigma_p(k,\mu,z)$ the errors of the logarithm of the observed power spectrum given by \cite{seo_probing_2003}:

\begin{equation}
\sigma_p = 2\pi\sqrt{\frac{2}{V_{\rm eff}(z,k,\mu) k^2 \Delta k \Delta \mu}\left(\frac{1 + n(z)P_{\rm obs,fid}}{n(z)P_{\rm obs,fid}} \right)} \, ,
\label{A2}
\end{equation}

\noindent with $n(z)$ the mean density of galaxies at a given redshift bin.

\item We finally compute the likelihood ratio:
\begin{equation}
L_R = {\rm exp}(-0.5(\chi^2_{\rm new} - \chi^2_{\rm previous})) \, .
\label{A3}
\end{equation}
\item We draw a random number $R_N$ in the interval $[0,1]$ from a uniform distribution.
\item We compare $L_R$ and $R_N$ : \\
- if $L_R$ $\geq$ $R_N$ the parameter set is accepted (the new parameter set is updated) \\
- else the parameter set is not accepted (the new parameter set is not updated).
\item We loop over 2 to 6.
\end{enumerate}

Finally we stop the chains using the Gelman-Rubin diagnostic, described in Appendix~\ref{sec:Gelman-Rubin}.

\section{The Gelman-Rubin diagnostic} 
\label{sec:Gelman-Rubin}

The Gelman-Rubin diagnostic aims to monitor MCMC outputs for parallelized chains. The convergence test computes a constant, ($R-1$), proportional to the ratio of the parameters variance of one chain to the mean parameters variances of the independent chains. The mean of the empirical variance within each chain $W$ is defined as: 

\begin{equation}
W = \frac{1}{M(N-1)}\sum_{i=1}^N\sum_{\alpha=1}^P(v_i^\alpha - \hat{v}^\alpha)^2 \, ,
\label{B1}
\end{equation}

\noindent with  $M$ the number of chains, $P$ the number of parameters and $N$ the number of accepted points for each chain; $\hat{v}^\alpha$ represents the mean value drawn of a parameter $\alpha$ for one chain and $v_i^\alpha$ the i-th $\alpha$ value drawn. The between-chain variance $B$ is given by: 

\begin{equation}
\frac{B}{N} = \frac{1}{(M-1)}\sum_{\alpha=1}^P(v_i^\alpha - \hat{v})^2 \, .
\label{B2}
\end{equation}

\noindent The weighted sum of $W$ and $B$ is written as :

\begin{equation}
V = \frac{N-1}{N}W + \frac{B}{N}\left(1 + \frac{1}{N}\right) \, .
\label{B3}
\end{equation}

\noindent Finally, the $R$ number is defined as:

\begin{equation}
R = \sqrt{\frac{\hat{d}+3}{\hat{d}+1} \frac{V}{W}} \geq 1 \, ,
\label{B4}
\end{equation}

\noindent with $\hat{d}$ the degrees of freedom estimate of a given distribution. The convergence is reached once $(R-1) < 0.03$ (i.e., $R$ must be sufficiently close to 1). 

\begin{table}
\centering
\setlength\tabcolsep{1pt}
\begin{tabular}{|c|c||c|c|}
  \hline
  $\Omega_{\rm b}$ & 4.518e-4 & $h$ & 1.900e-3 \\
  \hline
  $\Omega_{\rm m}$ & 3.657e-4 & $n_{\rm s}$ & 6.984e-4 \\
  \hline
  $\Omega_{\rm DE}$ & 1.807e-3 & $w_0$ & 4.007e-5 \\
  \hline
  $w_a$ & 3.765e-4 & $\sigma_8$ & 1.528e-4 \\
  \hline
  $\ln(b\sigma_8(0.125)$ & 1.316e-3 & $P_{\rm shot}(0.125)$ & 4.903e-5 \\
  \hline
  $\ln(b\sigma_8(0.375)$ & 1.091e-3 & $P_{\rm shot}(0.375)$ & 7.877e-4 \\
  \hline
  $\ln(b\sigma_8(0.6)$ & 5.615e-4 & $P_{\rm shot}(0.6)$ & 8.707e-5 \\
  \hline
  $\ln(b\sigma_8(0.8)$ & 7.964e-4 & $P_{\rm shot}(0.8)$ & 1.920e-4\\
  \hline
  $\ln(b\sigma_8(1.05)$ & 7.551e-4 & $P_{\rm shot}(1.05)$ & 1.213e-3 \\
  \hline
  $\ln(b\sigma_8(1.35)$ & 8.478e-3 & $P_{\rm shot}(1.35)$ & 1.458e-3 \\
  \hline
  $\ln(b\sigma_8(1.65)$ & 5.064e-4 & $P_{\rm shot}(1.65)$ & 1.003e-3 \\
  \hline
  $\ln(b\sigma_8(1.95)$ & 4.007e-4 & $P_{\rm shot}(1.95)$ & 1.635e-3 \\
  \hline
  $\ln(b\sigma_8(2.25)$ & 2.594e-4 & $P_{\rm shot}(2.25)$ & 7.626e-4 \\
  \hline
  $\ln(b\sigma_8(2.55)$ & 1.823e-2 & $P_{\rm shot}(2.55)$ & 9.062e-4 \\
  \hline
\end{tabular}
\caption{$(R-1)$ values for approach M1.}
\label{AP_B1}
\end{table}

\begin{table}
\centering
\setlength\tabcolsep{1pt}
\begin{tabular}{|c|c||c|c|}
  \hline
  $\omega_{\rm b}$ & 1.811e-2 & $h$ & 1.820e-2 \\
  \hline
  $\omega_{\rm m}$ & 1.820e-2 & $n_{\rm s}$ & 1.827e-2 \\
  \hline
  $\ln(D_a(0.125))$ & 1.812e-2 & $\ln(H(0.125))$ & 1.827e-2 \\
  \hline
  $\ln(f\sigma_8(0.125))$ & 1.819e-2 & $\ln(b\sigma_8(0.125))$ & 1.819e-2 \\
  \hline
  $P_{\rm shot}(0.125)$ & 1.817e-2 & $\ln(D_a(0.375))$ & 1.808e-2 \\
  \hline
  $\ln(H(0.375))$ & 1.817e-2 & $\ln(f\sigma_8(0.375))$ & 1.802e-2 \\
  \hline
  $\ln(b\sigma_8(0.375))$ & 1.819e-2 & $P_{\rm shot}(0.375)$ & 1.808e-2 \\
  \hline
  $\ln(D_a(0.6))$ & 1.821e-2 & $\ln(H(0.6))$ & 1.825e-2 \\
  \hline
  $\ln(f\sigma_8(0.6))$ & 1.811e-2 & $\ln(b\sigma_8(0.6))$ & 1.828e-2 \\
  \hline
  $P_{\rm shot}(0.6)$ & 1.809e-2 & $\ln(D_a(0.8))$ & 1.807e-2 \\
  \hline
  $\ln(H(0.8))$ & 1.810e-2 & $\ln(f\sigma_8(0.8))$ & 1.816e-2 \\
  \hline
  $\ln(b\sigma_8(0.8))$ & 1.813e-2 & $P_{\rm shot}(0.8)$ & 1.815e-2 \\
  \hline
  $\ln(D_a(1.05))$ & 1.8176e-2 & $\ln(H(1.05))$ & 1.814e-2 \\
  \hline
  $\ln(f\sigma_8(1.05))$ & 1.825e-2 & $\ln(b\sigma_8(1.05))$ & 1.818e-2 \\
  \hline
  $P_{\rm shot}(1.05)$ & 1.811e-2 & $\ln(D_a(1.35))$ & 1.814e-2 \\
  \hline
  $\ln(H(1.35))$ & 1.819e-2 & $\ln(f\sigma_8(1.35))$ & 1.811e-2 \\
  \hline
  $\ln(b\sigma_8(1.35))$ & 1.821e-2 & $P_{\rm shot}(1.35)$ & 1.822e-2 \\
  \hline
  $\ln(D_a(1.65))$ & 1.822e-2 & $\ln(H(1.65))$ & 1.815e-2 \\
  \hline
  $\ln(f\sigma_8(1.65))$ & 1.825e-2 & $\ln(b\sigma_8(1.65))$ & 1.821e-2 \\
  \hline
  $P_{\rm shot}(1.65)$ & 1.804e-2 & $\ln(D_a(1.95))$ & 1.820e-2 \\
  \hline
  $\ln(H(1.95))$ & 1.820e-2 & $\ln(f\sigma_8(1.95))$ & 1.821e-2 \\
  \hline
  $\ln(b\sigma_8(1.95))$ & 1.823e-2 & $P_{\rm shot}(1.95)$ & 1.812e-2 \\
  \hline
  $\ln(D_a(2.25))$ & 1.820e-2 & $\ln(H(2.25))$ & 1.820e-2 \\
  \hline
  $\ln(f\sigma_8(2.25))$ & 1.826e-2 & $\ln(b\sigma_8(2.25))$ & 1.820e-2 \\
  \hline
  $P_{\rm shot}(2.25)$ & 1.827e-2 & $\ln(D_a(2.25))$ & 1.820e-2 \\
  \hline
  $\ln(H(2.55))$ & 1.816e-2 & $\ln(f\sigma_8(2.55))$ & 1.811e-2 \\
  \hline
  $\ln(b\sigma_8(2.55))$ & 1.823e-2 & $P_{\rm shot}(2.55)$ & 1.830e-2 \\
  \hline
\end{tabular}
\caption{$(R-1)$ values for approach M2.}
\label{AP_B2}
\end{table}

\section{General definitions}
\label{sec:definitions}
In this last part of the Appendix we summarize the main definitions relevant for this work. The reader will find here a point of reference for the different definitions spread over the text.

\begin{description}
\item[\textbf{Optimal step}:]
Step size providing the minimum relative error between the analytical form and the numerical solutions. It ensures the most accurate derivative. 

\item[\textbf{Critical step}:]
Step size that can potentially lead to an error of 10\% on the final constraints.

\item[\textbf{Hypercritical step}:]
Step size that can potentially lead to catastrophic errors (several dozen percent) on the final constraints.

\item[\textbf{Truncation error}:]
Error introduced by considering too large step sizes. The approximation of computing the derivative with a finite amount of terms is in this case no longer valid.

\item[\textbf{Round-off error}:]
Error introduced by considering too small step sizes. The finite precision floating point numbers used on computers cannot distinguish between numbers that are too close.

\item[\textbf{Vibration matrix}:]
Matrix providing the largest value of the vibration for which 68\% of the constraints on each parameter remain smaller than the chosen precision.

\end{description}

\end{document}